\documentclass[a4paper,11pt]{article}
\usepackage{jheppub} 
\usepackage{lineno}

\usepackage{amssymb}
\usepackage{amsmath}
\usepackage{epsfig}
\usepackage{hyperref}
\usepackage{breakurl}
\usepackage{multirow}
\usepackage{array}
\usepackage{bm}
\usepackage{color}
\usepackage{cancel}

\usepackage{tikz}
\usetikzlibrary{snakes}
\usetikzlibrary{decorations}
\usetikzlibrary{trees}
\usetikzlibrary{decorations.pathmorphing}
\usetikzlibrary{decorations.markings}
\usetikzlibrary{external}
\usetikzlibrary{intersections}
\usetikzlibrary{shapes,arrows}
\usetikzlibrary{decorations.shapes}
\usetikzlibrary{arrows.meta}
\usetikzlibrary{calc}
\usetikzlibrary{shapes.misc}
\usetikzlibrary{decorations.text}
\usetikzlibrary{backgrounds}
\usetikzlibrary{fadings}
\usetikzlibrary{tikzmark,calc,arrows,shapes,decorations.pathreplacing}

\tikzset{
	graviton/.style={decorate,line width=0.15mm, 
	decoration={snake,amplitude=.5mm, segment length=3mm}
	},
	deflection/.style={dashed, line width = 0.6mm
	},
	massive/.style={postaction={decorate},
		line width=0.4mm
	},
	massiveWithArrow/.style={postaction={decorate},
		line width=0.4mm,
	},
	massiveWithArrowR/.style={postaction={decorate},
		line width=0.4mm,
	},
	massiveWithArrowB/.style={postaction={decorate},
		line width=0.75mm,
		decoration={
			markings,
			mark=at position 0.95 with {\arrow{latex}}}
	},
	cut/.style={dashed, line width = 0.5mm,mgreen
	},
	bgInsertion/.style={decorate, decoration={shape backgrounds, shape=circle, shape size=2mm}, style={fill=black}} ,
	amplitude/.style={decorate, decoration={shape backgrounds, shape=circle, shape size=14}, style={fill=gray}
	},
	smallBgInsertion/.style={decorate, decoration={shape backgrounds, shape=circle, shape size=1.5mm}, style={fill=black}
	} 
}

\newcommand{\FullAmplitude}[1]{
    \draw[amplitude] #1;
     \draw[smallBgInsertion] #1;
    }

\colorlet{mred}{black!30!red}
\colorlet{mgreen}{black!50!green}
\colorlet{mblue}{black!30!blue}
\colorlet{morange}{blue!70!red}

\newcommand{\scalarPropagator}{
\begin{tikzpicture}[baseline = {([yshift=4.2]current bounding box.center)}]

    \coordinate (e1) at (0,0);  
    \coordinate (e2) at (1,0);  
	\coordinate (e3) at (2,0);
 
    \draw[massiveWithArrow] (e1) -- (e3);

    \node[below] at (e2) {$p$}; 
		
\end{tikzpicture}}

\newcommand{\scalarBgVertex}{
\begin{tikzpicture}

    \coordinate (e1) at (0,0);  
    \coordinate (e2) at (1,0);  
	\coordinate (e3) at (2,0);

    \draw[massive] (e1) -- (e3);
    \draw[bgInsertion] (e2);

    \node[left] at (e1) {$p_1$};
    \node[right] at (e3) {$p_2$};
		
\end{tikzpicture}}

\newcommand{\gravitonPropagator}{
\begin{tikzpicture}[baseline = {([yshift=-.5]current bounding box.center)}]

    \coordinate (e1) at (0,0);  
    \coordinate (e2) at (1,0);  
	\coordinate (e3) at (2,0);

    \draw[graviton] (e1) -- (e3);

    \node[below] at (e2) {$q$}; 
    \node[left] at (e1) {$\mu\nu$};
    \node[right] at (e3) {$\alpha\beta$};
		
\end{tikzpicture}}

\newcommand{\gravitonBgVertex}{
\begin{tikzpicture}

  \node[left] at (0,0) {$(q_1,\mu\nu)$};
  \node[right] at (2,0) {$(q_2,\alpha\beta)$};
  
    \draw [graviton] (0,0) -- (2,0);
    \draw [bgInsertion] (1,0);
\end{tikzpicture}}

\newcommand{\scalarGravitonVertex}{
\begin{tikzpicture}

    \coordinate (e1) at (0,0);  
    \coordinate (e2) at (1,0);  
	\coordinate (e3) at (2,0);
	\coordinate (e4) at (1,1);

    \draw[massiveWithArrow] (e1) -- (e2);
    \draw[massiveWithArrow] (e2) -- (e3);
    \draw[graviton] (e2) -- (e4);

    \node[left] at (e1) {$p_1$};
    \node[right] at (e3) {$p_2$};
    \node[above] at (e4) {$(q,\mu\nu)$};

    \coordinate (eC1) at (4,0);  
    \coordinate (eC2) at (5,0);  
	\coordinate (eC3) at (6,0);
	\coordinate (eC4) at (5,1);

    \draw[massive] (eC1) -- (eC3);
    \draw[graviton] (eC2) -- (eC4);
    \draw[bgInsertion] (eC2);

    \node[left] at (eC1) {$p_1$};
    \node[right] at (eC3) {$p_2$};
    \node[above] at (eC4) {$(q,\mu\nu)$};
		
\end{tikzpicture}}

\newcommand{\deflectionPropagator}{
\begin{tikzpicture}[baseline = {([yshift=-1]current bounding box.center)}]

    \coordinate (e1) at (0,0);  
    \coordinate (e2) at (1,0);  
	\coordinate (e3) at (2,0);

    \draw[deflection] (e1) -- (e3);

    \node[left] at (e1) {$\alpha$};
    \node[right] at (e3) {$\beta$};
    \node[below] at (e2) {$\omega$}; 
		
\end{tikzpicture}}

\newcommand{\deflectionGravitonVertex}{
\begin{tikzpicture}

    \coordinate (e1) at (0,0);  
    \coordinate (e2) at (1,0);  
	\coordinate (e3) at (2,0);

    \draw[deflection] (e1) -- (e2);
    \draw[graviton] (e2) -- (e3);

    \node[left] at (e1) {$(\omega,\alpha)$};
    \node[right] at (e3) {$(q,\mu\nu)$};
		
\end{tikzpicture}}

\newcommand{\comptonFeynGraphs}{
\begin{tikzpicture}

    \coordinate (eF0) at (-3,0);  
    \coordinate (eFL) at (-4,-.5);  
	\coordinate (eFR) at (-2,-.5);
	
	\coordinate (eq) at (-1,0);

    \coordinate (e0) at (1,0);  
    \coordinate (eL) at (0,-.5);  
	\coordinate (eR) at (2,-.5);
	
	\coordinate (plus) at (3,0);
	
	\coordinate (e0D) at (4.7,0);
	\coordinate (e00D) at (5.3,0);  
    \coordinate (eLD) at (4,-.5);  
	\coordinate (eRD) at (6,-.5);
    
    \node[left] at (eL) {6};
    \node[right] at (eR) {7};
    \node[left] at (eLD) {6};
    \node[right] at (eRD) {7};
    \node[left] at (eFL) {6};
    \node[right] at (eFR) {7};
    
    \node at (eq) {$=$};
    \node at (plus) {$+$};
    
    \draw [graviton] (eFL) -- (eF0);
    \draw [graviton] (eF0) -- (eFR);
    \draw [amplitude] (eF0);
    
    \draw [graviton] (eL) -- (e0);
    \draw [graviton] (e0) -- (eR);
    \draw [bgInsertion] (e0);
    
    \draw [graviton] (eLD) -- (e0D);
    \draw [graviton] (e00D) -- (eRD);
    \draw[deflection] (e0D) -- (e00D);
    		
\end{tikzpicture}}

\def\offset{6}

\newcommand{\LeadingGoneSF}{
\begin{tikzpicture}[radius=1]

	\coordinate (e1) at (0,0);  
	\coordinate (e2) at (1,0);
	\coordinate (e3) at (3,0);
	\coordinate (e4) at (4,0);
	\coordinate (e5) at (1.95,1.0);

 	 \draw[graviton] (e2)  arc[start angle=180, end angle=90];
 	 \draw[graviton] (e3)  arc[start angle=0, end angle=90];
 	 \draw[massiveWithArrow] (e1) -- (e4);
	 \draw[bgInsertion] (e5);
	
    \node[left] at (e1) {2};
    \node[right] at (e4) {3};
    
	\coordinate (eD1) at (\offset+0,0);  
	\coordinate (eD2) at (\offset+1,0);
	\coordinate (eD3) at (\offset+3.7,0);
	\coordinate (eD4) at (\offset+4.7,0);
	\coordinate (eD5) at (\offset+2,1);
	\coordinate (eD6) at (\offset+2.7,1);

	\draw[graviton] (eD2)  arc[start angle=180, end angle=90];
	\draw[deflection] (eD5) -- (eD6);
	\draw[graviton] (eD3)  arc[start angle=0, end angle=90];
	\draw[massiveWithArrow] (eD1) -- (eD4);
		
    \node[left] at (eD1) {2};
    \node[right] at (eD4) {3};
		
\end{tikzpicture}}

\def\diagNN{-1.8}

\newcommand{\FivePtFeynGraphs}{
\begin{tikzpicture}

    \coordinate (eF0) at (-3,0);  
    \coordinate (eFL) at (-4,0);  
	\coordinate (eFR) at (-2,0);
	\coordinate (eFU) at (-2,1);
    
    \draw [massiveWithArrow] (eFL) -- (eF0);
    \draw [massiveWithArrowR] (eF0) -- (eFR);
    \draw [graviton] (eF0) -- (eFU);
    \draw [amplitude] (eF0);
    
    \node[left] at (eFL) {2};
    \node[right] at (eFR) {3};
    \node[right] at (eFU) {5};
	
	\coordinate (eq) at (-1,0);
	\node at (eq) {$=$};
	
    \coordinate (e1L) at (0,0);  
    \coordinate (e100) at (.7,0);  
    \coordinate (e10) at (1.3,0);  
	\coordinate (e1R) at (2,0);
	\coordinate (e1U) at (2,1);
    
    \draw [massiveWithArrow] (e1L) -- (e100);
    \draw [massiveWithArrowR] (e100) -- (e10);
    \draw [massiveWithArrowR] (e10) -- (e1R);
    \draw [bgInsertion] (e100);
    \draw [graviton] (e10) -- (e1U);
    
    \node[left] at (e1L) {2};
    \node[right] at (e1R) {3};
    \node[right] at (e1U) {5};
	
	\coordinate (plus) at (3,0);
    \node at (plus) {$+$};
    
    \coordinate (e2L) at (4,0);  
    \coordinate (e200) at (4.7,0);  
    \coordinate (e20) at (5.3,0);  
	\coordinate (e2R) at (6,0);
	\coordinate (e2U) at (6,1);
    
    \draw [massiveWithArrow] (e2L) -- (e200);
    \draw [massiveWithArrowR] (e200) -- (e20);
    \draw [massiveWithArrowR] (e20) -- (e2R);
    \draw [graviton] (e200) -- (e2U);
    \draw [bgInsertion] (e20);
    
    \node[left] at (e2L) {2};
    \node[right] at (e2R) {3};
    \node[right] at (e2U) {5};
	
	\coordinate (plus2) at (-1,\diagNN);
    \node at (plus2) {$+$};
    
    \coordinate (plus25) at (7,0);
    \node at (plus25) {$+$};
	
    \coordinate (e25L) at (8,0);  
    \coordinate (e250) at (9,0);  
	\coordinate (e25R) at (10,0);
	\coordinate (e25U) at (10,1);
    
    \draw [massiveWithArrow] (e25L) -- (e250);
    \draw [massiveWithArrowR] (e250) -- (e25R);
    \draw [graviton] (e250) -- (e25U);
    \draw [bgInsertion] (e250);
    
    \node[left] at (e25L) {2};
    \node[right] at (e25R) {3};
    \node[right] at (e25U) {5};
    
    \coordinate (e3L) at (0,\diagNN);  
    \coordinate (e30) at (.7,\diagNN);   
	\coordinate (e3R) at (2,\diagNN);
	\coordinate (e3U1) at (.7,\diagNN+.7);
	\coordinate (e3U2) at (1.3,\diagNN+1);
	\coordinate (e3U3) at (2,\diagNN+1);
    
    \draw [massiveWithArrow] (e3L) -- (e30);
    \draw [massiveWithArrowR] (e30) -- (e3R);
    \draw [graviton] (e30) -- (e3U1);
    \draw [deflection] (e3U1) -- (e3U2);
    \draw [graviton] (e3U2) -- (e3U3);
    
    \node[left] at (e3L) {2};
    \node[right] at (e3R) {3};
    \node[right] at (e3U3) {5};
    
	\coordinate (plus3) at (3,\diagNN);
    \node at (plus3) {$+$};
    
    \coordinate (e4L) at (4,\diagNN);  
    \coordinate (e40) at (5,\diagNN);   
	\coordinate (e4R) at (6,\diagNN);
	\coordinate (e4U1) at (5,\diagNN+1);
	\coordinate (e4U2) at (6,\diagNN+1);
    
    \draw [massiveWithArrow] (e4L) -- (e40);
    \draw [massiveWithArrowR] (e40) -- (e4R);
    \draw [graviton] (e40) -- (e4U1);
    \draw [graviton] (e4U1) -- (e4U2);
    \draw [bgInsertion] (e4U1);
    
    \node[left] at (e4L) {2};
    \node[right] at (e4R) {3};
    \node[right] at (e4U2) {5};
    		
\end{tikzpicture}}

\newcommand{\TwoLoopUnitarity}{
\begin{tikzpicture}[radius=1]

    \coordinate (eL) at (0,0);  
    \coordinate (v1) at (1,0);  
    \coordinate (v2) at (3,0);  
	\coordinate (eR) at (4,0);
	\coordinate (eU) at (2,1);
	
	\coordinate (c1) at (2,-.3);
	\coordinate (c2) at (2,1.3);
    
    \draw [massiveWithArrow] (eL) -- (v1);
    \draw [massiveWithArrow] (v1) -- (v2);
    \draw [massiveWithArrowR] (v2) -- (eR);

 	 \draw[graviton] (v1)  arc[start angle=180, end angle=90];
 	 \draw[graviton] (v2)  arc[start angle=0, end angle=90];
    
    \node[left] at (eL) {2};
    \node[right] at (eR) {3};
    
    \draw[cut] (c1) -- (c2);
    \draw[amplitude] (v1);
    \draw[amplitude] (v2);
    		
\end{tikzpicture}}

\newcommand{\TwoLoopFeynRules}{
\begin{tikzpicture}[radius=1]

    \coordinate (e1v1) at (-4,0);  
    \coordinate (e1v2) at (-2,0);  
    \coordinate (e1U) at (-3,1);  
    \coordinate (e1L) at (-5,0);  
    \coordinate (e1BG) at (-4.5,0);  
	\coordinate (e1R) at (-1,0);
    
    \draw [massiveWithArrowR] (e1L) -- (e1BG);
    \draw [massiveWithArrowR] (e1BG) -- (e1v1);
    \draw [massiveWithArrowR] (e1v1) -- (e1v2);
    \draw [massiveWithArrowR] (e1v2) -- (e1R);
    \draw [bgInsertion] (e1BG);
    \draw [bgInsertion] (e1U);

 	 \draw[graviton] (e1v1)  arc[start angle=180, end angle=90];
 	 \draw[graviton] (e1v2)  arc[start angle=0, end angle=90];
    
    \node[left] at (e1L) {2};
    \node[right] at (e1R) {3};
    
    \coordinate (e2v1) at (1,0);  
    \coordinate (e2v2) at (3,0);  
    \coordinate (e2L) at (0,0);  
    \coordinate (e2BG) at (2,1);  
	\coordinate (e2R) at (4,0);
    
    \draw [massiveWithArrowR] (e2L) -- (e2v1);
    \draw [massiveWithArrowR] (e2v1) -- (e2v2);
    \draw [massiveWithArrowR] (e2v2) -- (e2R);
    \draw [bgInsertion] (e2BG);
    \draw [bgInsertion] (e2v1);

 	 \draw[graviton] (e2v1)  arc[start angle=180, end angle=90];
 	 \draw[graviton] (e2v2)  arc[start angle=0, end angle=90];
    
    \node[left] at (e2L) {2};
    \node[right] at (e2R) {3};
	  
    \coordinate (e3v1) at (6,0);  
    \coordinate (e3v2) at (8,0);  
    \coordinate (e3L) at (5,0);  
    \coordinate (e3BG) at (7,0);  
    \coordinate (e3U) at (7,1);  
	\coordinate (e3R) at (9,0);
    
    \draw [massiveWithArrowR] (e3L) -- (e3v1);
    \draw [massiveWithArrowR] (e3v1) -- (e3BG);
    \draw [massiveWithArrowR] (e3BG) -- (e3v2);
    \draw [massiveWithArrowR] (e3v2) -- (e3R);
    \draw [bgInsertion] (e3BG);
    \draw [bgInsertion] (e3U);

 	 \draw[graviton] (e3v1)  arc[start angle=180, end angle=90];
 	 \draw[graviton] (e3v2)  arc[start angle=0, end angle=90];
    
    \node[left] at (e3L) {2};
    \node[right] at (e3R) {3};

    \coordinate (e1v1) at (-4.25,\diagNN);  
    \coordinate (e1v2) at (-1.75,\diagNN);  
    \coordinate (e1U1) at (-3.25,\diagNN+1);  
    \coordinate (e1U2) at (-2.75,\diagNN+1);  
    \coordinate (e1L) at (-5,\diagNN);  
    \coordinate (e1BG) at (-4.6,\diagNN);  
	\coordinate (e1R) at (-1,\diagNN);
    
    \draw [massive] (e1L) -- (e1BG);
    \draw [massive] (e1BG) -- (e1v1);
    \draw [massiveWithArrowR] (e1v1) -- (e1v2);
    \draw [massiveWithArrowR] (e1v2) -- (e1R);
    \draw [bgInsertion] (e1BG);
    \draw [bgInsertion] (e1U);

 	 \draw[graviton] (e1v1)  arc[start angle=180, end angle=90];
 	 \draw[graviton] (e1v2)  arc[start angle=0, end angle=90];
   \draw[deflection] (e1U1) -- (e1U2);
    
    \node[left] at (e1L) {2};
    \node[right] at (e1R) {3};
    
    \coordinate (e2v1) at (.75,\diagNN);  
    \coordinate (e2v2) at (3.25,\diagNN);  
    \coordinate (e2L) at (0,\diagNN);  
    \coordinate (e2U1) at (1.75,\diagNN+1);  
    \coordinate (e2U2) at (2.25,\diagNN+1); 
	\coordinate (e2R) at (4,\diagNN);
    
    \draw [massiveWithArrowR] (e2L) -- (e2v1);
    \draw [massiveWithArrowR] (e2v1) -- (e2v2);
    \draw [massiveWithArrowR] (e2v2) -- (e2R);
    \draw [bgInsertion] (e2BG);
    \draw [bgInsertion] (e2v1);

 	 \draw[graviton] (e2v1)  arc[start angle=180, end angle=90];
 	 \draw[graviton] (e2v2)  arc[start angle=0, end angle=90];
   \draw[deflection] (e2U1) -- (e2U2);
    
    \node[left] at (e2L) {2};
    \node[right] at (e2R) {3};
	  
    \coordinate (e3v1) at (5.75,\diagNN);  
    \coordinate (e3v2) at (8.25,\diagNN);  
    \coordinate (e3L) at (5,\diagNN);  
    \coordinate (e3BG) at (7,\diagNN);  
    \coordinate (e3U1) at (6.75,\diagNN+1);  
    \coordinate (e3U2) at (7.25,\diagNN+1);  
	\coordinate (e3R) at (9,\diagNN);
    
    \draw [massiveWithArrowR] (e3L) -- (e3v1);
    \draw [massiveWithArrowR] (e3v1) -- (e3BG);
    \draw [massiveWithArrowR] (e3BG) -- (e3v2);
    \draw [massiveWithArrowR] (e3v2) -- (e3R);
    \draw [bgInsertion] (e3BG);

 	 \draw[graviton] (e3v1)  arc[start angle=180, end angle=90];
 	 \draw[graviton] (e3v2)  arc[start angle=0, end angle=90];
   \draw[deflection] (e3U1) -- (e3U2);
    
    \node[left] at (e3L) {2};
    \node[right] at (e3R) {3};

    \coordinate (e4v1) at (-4,2*\diagNN);  
    \coordinate (e4v2) at (-2,2*\diagNN);  	
    \coordinate (e4L) at (-5,2*\diagNN);  
    \coordinate (e40) at (-2,2*\diagNN+0.95);   
    \coordinate (e400) at (-4,2*\diagNN+0.95);   
	\coordinate (e4R) at (-1,2*\diagNN);
    
    \draw [massiveWithArrow] (e4L) -- (e4v1);
    \draw [massiveWithArrowR] (e4v1) -- (e4v2);    
    \draw [massiveWithArrowR] (e4v2) -- (e4R);
    \draw [bgInsertion] (e40);
    \draw [bgInsertion] (e400);

   \draw[graviton] (e4v1) -- (e400);
   \draw[graviton] (e4v2) -- (e40);
   \draw[graviton] (e400) -- (e40);
    
    \node[left] at (e4L) {2};
    \node[right] at (e4R) {3};

    \coordinate (e5v1) at (1,2*\diagNN);  
    \coordinate (e5v2) at (3,2*\diagNN);  
    \coordinate (e5L) at (0,2*\diagNN);   
    \coordinate (e50) at (3,2*\diagNN+0.95);   
    \coordinate (e500) at (1,2*\diagNN+0.95); 
	\coordinate (e5R) at (4,2*\diagNN);
    \coordinate (e5UD1) at (3,2*\diagNN+0.65);   
    \coordinate (e5UD2) at (3,2*\diagNN+0.95); 
    \coordinate (e5UD3) at (2.5,2*\diagNN+0.95);

    \draw [massiveWithArrowR] (e5L) -- (e5v1);
    \draw [massiveWithArrowR] (e5v1) -- (e5v2);
    \draw [massiveWithArrowR] (e5v2) -- (e5R);
    \draw [bgInsertion] (e500);

   \draw[graviton] (e5v1) -- (e500);
   \draw[graviton] (e5v2) -- (e5UD1);
   \draw[graviton] (e500) -- (e5UD3);
   \draw[deflection] (e5UD3) -- (e5UD2);
   \draw[deflection] (e5UD1) -- (e5UD2);
    
    \node[left] at (e5L) {2};
    \node[right] at (e5R) {3};
    
    \coordinate (e3v1) at (6,0);  
    \coordinate (e3v2) at (8,0);  
    \coordinate (e3L) at (5,0);  
    \coordinate (e3BG) at (7,0);  
    \coordinate (e3U) at (7,1);  
	\coordinate (e3R) at (9,0);
    
    \coordinate (e6v1) at (6,2*\diagNN);  
    \coordinate (e6v2) at (8,2*\diagNN);  
    \coordinate (e6L) at (5,2*\diagNN);   
    \coordinate (e60) at (8,2*\diagNN+0.95);   
    \coordinate (e600) at (6,2*\diagNN+0.95); 
	\coordinate (e6R) at (9,2*\diagNN);
    \coordinate (e6UD1) at (8,2*\diagNN+0.65);   
    \coordinate (e6UD2) at (8,2*\diagNN+0.95); 
    \coordinate (e6UD3) at (7.5,2*\diagNN+0.95); 
    \coordinate (e6uD1) at (6,2*\diagNN+0.65);   
    \coordinate (e6uD2) at (6,2*\diagNN+0.95); 
    \coordinate (e6uD3) at (6.5,2*\diagNN+0.95);

    \draw [massiveWithArrowR] (e6L) -- (e6v1);
    \draw [massiveWithArrowR] (e6v1) -- (e6v2);
    \draw [massiveWithArrowR] (e6v2) -- (e6R);

   \draw[graviton] (e6v1) -- (e6uD1);
   \draw[graviton] (e6v2) -- (e6UD1);
   \draw[graviton] (e6uD3) -- (e6UD3);
   \draw[deflection] (e6UD3) -- (e6UD2);
   \draw[deflection] (e6UD1) -- (e6UD2);
   \draw[deflection] (e6uD3) -- (e6uD2);
   \draw[deflection] (e6uD1) -- (e6uD2);
    
    \node[left] at (e6L) {2};
    \node[right] at (e6R) {3};

\end{tikzpicture}}

\def\offsetSF{3}
\def\offsetSFT{5}
\def\nodeSF{-.7}
\newcommand{\SFCounting}{
\begin{tikzpicture}

    \coordinate (e0L) at (-\offsetSF,0);  
    \coordinate (v0) at (1-\offsetSF,0);  
    \coordinate (v0Below) at (1-\offsetSF,\nodeSF);  
	\coordinate (e0R) at (2-\offsetSF,0);
    
    \draw [massiveWithArrow] (e0L) -- (v0);
    \draw [massiveWithArrowR] (v0) -- (e0R);
    \FullAmplitude{(v0)}
    
    \node[left] at (e0L) {2};
    \node[right] at (e0R) {3};
    
    \coordinate (eL) at (0,0);  
    \coordinate (v1) at (1,0);  
    \coordinate (vBelow) at (2,\nodeSF);  
    \coordinate (v2) at (3,0);  
	\coordinate (eR) at (4,0);
	\coordinate (eU) at (2,1);
    
    \draw [massiveWithArrow] (eL) -- (v1);
    \draw [massiveWithArrow] (v1) -- (v2);
    \draw [massiveWithArrowR] (v2) -- (eR);

 	\draw[graviton] (v1)  arc[start angle=180, end angle=90, radius=1];
 	\draw[graviton] (v2)  arc[start angle=0, end angle=90, radius=1];
    
    \node[left] at (eL) {2};
    \node[right] at (eR) {3};

    \FullAmplitude{(v1)}
    \FullAmplitude{(v2)}
    
    \coordinate (e2L) at (\offsetSFT,0);  
    \coordinate (v21) at (1+\offsetSFT,0);  
    \coordinate (v22) at (2+\offsetSFT,0);  
    \coordinate (v22Below) at (2+\offsetSFT,\nodeSF);  
    \coordinate (v23) at (3+\offsetSFT,0);  
	\coordinate (e2R) at (4+\offsetSFT,0);
	\coordinate (e2U) at (2+\offsetSFT,1);
    
    \draw [massiveWithArrow] (e2L) -- (v21);
    \draw [massiveWithArrow] (v21) -- (v22);
    \draw [massiveWithArrow] (v22) -- (v23);
    \draw [massiveWithArrowR] (v23) -- (e2R);

 	\draw[graviton] (v21)  arc[start angle=180, end angle=90, radius=1];
 	\draw[graviton] (v23)  arc[start angle=0, end angle=90, radius=1];
    \draw[graviton] (v22) -- (e2U);
    \draw[bgInsertion] (e2U);
    
    \node[left] at (e2L) {2};
    \node[right] at (e2R) {3};

    \FullAmplitude{(v21)}
    \FullAmplitude{(v22)}
    \FullAmplitude{(v23)}

    \node at (v0Below) {0SF};
    \node at (vBelow) {1SF};
    \node at (v22Below) {2SF};
    		
\end{tikzpicture}}

\def\offsetHourT{4}
\def\offsetHour{6}
\def\downsetHour{-4}
\def\offsetHourTwo{11}
\newcommand{\Hourglass}{
\begin{tikzpicture}[radius=1]

    \coordinate (BL) at (-2,-1);  
    \coordinate (BLG) at (-1,-1); 
    \coordinate (BRG) at (1,-1);  
    \coordinate (BR) at (2,-1);  
    \coordinate (C) at (0,0);
    \coordinate (TL) at (-2,1);  
    \coordinate (TLLG) at (-1.25,1); 
    \coordinate (TLG) at (-.75,1); 
    \coordinate (TRG) at (.75,1);  
    \coordinate (TRRG) at (1.25,1);  
    \coordinate (TR) at (2,1);  
    \coordinate (Cdots) at (0,.8);  
    \coordinate (Cbot) at (0,-1.25);

    \draw [massiveWithArrow] (BL) -- (BR);
    \draw [graviton] (BLG) -- (C);
    \draw [graviton] (BRG) -- (C);
    \draw [massiveWithArrow] (TL) -- (TR);
    \draw [graviton] (TLG) -- (C);
    \draw [graviton] (TRG) -- (C);
    \draw [graviton] (C) -- (TLLG);
    \draw [graviton] (C) -- (TRRG);
    \node at (Cdots) {$\ldots$};
    \node[left] at (BL) {2};
    \node[right] at (BR) {3};
    \node[left] at (TL) {1};
    \node[right] at (TR) {4};
    \node[below] at (Cbot) {Flat};

	\coordinate (e1) at (\offsetHour+0,0-1);  
	\coordinate (e2) at (\offsetHour+1,0-1);
	\coordinate (e3) at (\offsetHour+3,0-1);
	\coordinate (e4) at (\offsetHour+4,0-1);
	\coordinate (e5) at (\offsetHour+1.95,1.0-1);
    \coordinate (ebot) at (\offsetHour+1.95,-1.25);

 	 \draw[graviton] (e2)  arc[start angle=180, end angle=90];
 	 \draw[graviton] (e3)  arc[start angle=0, end angle=90];
 	 \draw[massiveWithArrow] (e1) -- (e4);
	 \draw[bgInsertion] (e5);
	
    \node[left] at (e1) {2};
    \node[right] at (e4) {3};
    \node[below] at (ebot) {Curved};

    \node at (\offsetHourT,-1+.5) {$\bm \longleftrightarrow$};
    
    \coordinate (BL) at (-2,-1+\downsetHour);  
    \coordinate (BLG) at (-1,-1+\downsetHour); 
    \coordinate (BRG) at (1,-1+\downsetHour);  
    \coordinate (BR) at (2,-1+\downsetHour);  
    \coordinate (CL) at (-1,0+\downsetHour);
    \coordinate (CR) at (1,0+\downsetHour);
    \coordinate (TL) at (-2,1+\downsetHour);  
    \coordinate (TLLG) at (-1.5,1+\downsetHour); 
    \coordinate (TLG) at (-.5,1+\downsetHour); 
    \coordinate (TRG) at (.5,1+\downsetHour);  
    \coordinate (TRRG) at (1.5,1+\downsetHour);  
    \coordinate (TR) at (2,1+\downsetHour);  
    \coordinate (CLdots) at (-1,.8+\downsetHour);    
    \coordinate (CRdots) at (1,.8+\downsetHour);  
    \coordinate (Cbot) at (0,-1.25+\downsetHour);

    \draw [massiveWithArrow] (BL) -- (BR);
    \draw [graviton] (BLG) -- (CL);
    \draw [graviton] (BRG) -- (CR);
    \draw [graviton] (CL) -- (CR);
    \draw [massiveWithArrow] (TL) -- (TR);
    \draw [graviton] (CL) -- (TLLG);
    \draw [graviton] (CL) -- (TLG);
    \node at (CLdots) {$\ldots$};
    \draw [graviton] (TRRG) -- (CR);
    \draw [graviton] (TRG) -- (CR);
    \node at (CRdots) {$\ldots$};
    \node[left] at (BL) {2};
    \node[right] at (BR) {3};
    \node[left] at (TL) {1};
    \node[right] at (TR) {4};
    \node[below] at (Cbot) {Flat};

    \coordinate (e4v1) at (-4+\offsetHourTwo,+\downsetHour-1);  
    \coordinate (e4v2) at (-2+\offsetHourTwo,+\downsetHour-1);  	
    \coordinate (e4L) at (-5+\offsetHourTwo,+\downsetHour-1);  
    \coordinate (e40) at (-2+\offsetHourTwo,0.95+\downsetHour-1);
    \coordinate (e4C) at (-3+\offsetHourTwo,-1.25+\downsetHour);   
    
    \coordinate (e400) at (-4+\offsetHourTwo,0.95+\downsetHour-1);   
	\coordinate (e4R) at (-1+\offsetHourTwo,+\downsetHour-1);
    
    \draw [massiveWithArrow] (e4L) -- (e4v1);
    \draw [massiveWithArrowR] (e4v1) -- (e4v2);    
    \draw [massiveWithArrowR] (e4v2) -- (e4R);
    \draw [bgInsertion] (e40);
    \draw [bgInsertion] (e400);

   \draw[graviton] (e4v1) -- (e400);
   \draw[graviton] (e4v2) -- (e40);
   \draw[graviton] (e400) -- (e40);
    
    \node[left] at (e4L) {2};
    \node[right] at (e4R) {3};
    \node[below] at (e4C) {Curved};

    \node at (\offsetHourT,-1+.5+\downsetHour) {$\bm \longleftrightarrow$};

\end{tikzpicture}}
\graphicspath{{figures/}}
\usepackage[export]{adjustbox}

\def\nn{\nonumber}

\newcommand{\eqtref}[1]{Eq.~(\ref{#1})}
\newcommand{\eqsref}[2]{Eqs.~(\ref{#1}) and~(\ref{#2})}
\newcommand{\secref}[1]{Sec.~\ref{#1}}
\newcommand{\secsref}[2]{Secs.~\ref{#1} and~\ref{#2}}
\newcommand{\figref}[1]{Fig.~\ref{#1}}

\def\sch{Schwarzschild }

\def\imUnit{i} 
\def\varphib{\bar{\varphi}} 
\def\gb{\bar{g}} 
\def\gA{A} 
\def\gB{B} 
\def\cov{\bar{\mathcal{D}}} 

\def\actionG{S_\text{G}}
\def\actionGFree{S_\text{G, free}}
\def\actionGInt{S_\text{G, int}}
\def\actionH{S_\text{H}}
\def\actionL{S_\text{L}}
\def\actionLInt{S_\text{L, int}}
\def\actionLY{S_\text{LY}}
\def\actionLYInt{S_\text{LY, int}}
\def\actionHFin{S_\text{H, fin}}
\def\actionHFinFree{S_\text{H, FF}}
\def\actionHFinInt{S_\text{H, FI}}

\begin{document}

\title{Gravitational Self Force from Scattering Amplitudes in Curved Space 
}

\author[1]{Dimitrios Kosmopoulos,}
\affiliation[1]{D\'epartement de Physique Th\'eorique, Universit\'e de Gen\`eve, 24 quai E. Ansermet, CH-1211 Geneva, Switzerland}
\emailAdd{dimitrios.kosmopoulos@unige.ch}
\author[2]{Mikhail P. Solon}
\affiliation[2]{
Mani L. Bhaumik Institute for Theoretical Physics, Department of Physics and Astronomy, University of California Los Angeles, Los Angeles, CA 90095, USA
}
\emailAdd{solon@physics.ucla.edu}

\abstract{
We employ scattering amplitudes in curved space to model the dynamics of a light probe particle with mass $m$ orbiting in the background spacetime induced by a heavy gravitational source with mass $M$. Observables are organized as an expansion in $m/M$ to all orders in $G$ --- the gravitational self-force expansion. An essential component of our analysis is the backreaction of the heavy source which we capture by including the associated light degrees of freedom. As illustration we consider a Schwarzschild background and verify geodesic motion as well as the first-order self-force correction to two-body scattering through ${\cal O}(G^3)$. Amplitudes in curved space offer several advantages, and further developments along these lines may advance the computation of gravitational-wave signals for extreme-mass-ratio inspirals.
}

\maketitle
\newpage

\section{Introduction}

The breakthrough discovery of gravitational waves~\cite{LIGOScientific:2016aoc} has catalyzed rapid progress and new directions in many areas of astronomy, cosmology, and particle physics.  
In theoretical high-energy physics, the challenge of modeling binary sources to high precision has inspired new methods of calculation using the tools of Quantum Field Theory (QFT), leading to a number of new results in General Relativity. In turn, the application of QFT in this new arena has uncovered theoretical structures that have brought new insights into the deep connections between field theory and Einstein's geometric picture of gravity. See, e.g., Refs.~\cite{Buonanno:2022pgc,Adamo:2022dcm,Bjerrum-Bohr:2022blt,Kosower:2022yvp} for reviews. 

The QFT-based approach to classical binary dynamics weaves together cutting-edge tools from the theory of elementary particles, such as effective field theory (EFT), on-shell methods, and techniques for multi-loop integration. These tools have been honed over decades to tackle significant problems in gauge theory where multiple scales are nonlinearly coupled. However, the precision demands of future gravitational wave detectors, such as LISA~\cite{LISA:2017pwj}, the Einstein Telescope~\cite{Punturo:2010zz} and Cosmic Explorer~\cite{Reitze:2019iox}, are posing new challenges that will take these technologies to their breaking-point.

State-of-the-art calculations provide a stress test for our understanding of QFT and unveil deep insights and underlying structure. A celebrated example is the double copy~\cite{Kawai:1985xq,Bern:2008qj,Bern:2010ue}, which, in its original incarnation, directly connected amplitudes in gauge and gravitational theories, and has since been generalized to a web of interconnected theories (see Ref.~\cite{Bern:2019prr} for a recent review). Computations relevant for the binary problem, in particular, have also led to formal developments. For example, Ref.~\cite{Cheung:2020gbf} obtained an all-orders-in-$G$ amplitude in the probe limit, Ref.~\cite{Cristofoli:2021jas} exposed relations among various amplitudes in the classical limit, while Ref.~\cite{Bern:2021dqo} conjectured an exponential form for the amplitude in terms of the classical radial action. These calculations demonstrate how fertile this program is for deepening our understanding of scattering amplitudes, which are some of the fundamental objects in QFT, and at the same time impacting the exciting frontier of gravitational-wave science. 

In the present work we extend the application of QFT methods to binary dynamics by exploring scattering amplitudes in curved space. We leverage exact solutions to Einstein's equations to develop a framework for systematically computing classes of contributions to all orders in the gravitational coupling $G$ but as an expansion in the small ratio of masses, $m/M \ll 1$, which characterizes so-called extreme-mass-ratio inspirals. More broadly, describing the dynamics around a non-trivial background is a rich, albeit underexplored, field-theory problem with generic aspects that can find applicability beyond the context of gravitational waves.

In attempting to address the binary inspiral problem analytically one is immediately led to the principles of EFT. Indeed, the separation of scales inherent in the problem, namely the distance between the two bodies in the inspiral phase being much larger than their size, allows one to treat the bodies as point particles. This line of reasoning led to the original treatment of the binary as two point particles following geodesics in a metric induced by their motion (see Ref.~\cite{Blanchet:2013haa} for a review). Later, the problem was reformulated explicitly as an EFT~\cite{Goldberger:2004jt} (see also the reviews \cite{Porto:2016pyg,Levi:2018nxp}). The same point of view is built into the modern QFT approaches to the problem as well (see Refs.~\cite{Bern:2021dqo,Bern:2021yeh,Dlapa:2021npj,Dlapa:2021vgp,Dlapa:2022lmu,Damgaard:2023ttc,Jakobsen:2023ndj} for remarkable recent results at 
4PM\footnote{The post-Minkowskian (PM) expansion is organized in powers of $G$: $n$PM $\leftrightarrow$ $\mathcal{O}\left(G^n\right)$.}). 

The above approaches do not assume a hierarchy between the masses of the bodies in the binary, rather they are perturbation schemes in Newton's constant $G$. Indeed, one of the open challenges is modeling the inspiral of binary systems consisting of a compact body of mass $m$ and a supermassive black hole of mass $M$, with $M \gg m$. The limiting case of these so-called extreme-mass-ratio inspirals is described by geodesic motion in a background spacetime, such as Schwarzschild or Kerr, but beyond this the probe particle interacts with its own gravitational field, giving rise to an effective “self-force.” The gravitational self-force framework aims to solve this system as an expansion in $m/M$ but to all orders in $G$. In this framework, the $n$th order self-force ($n$SF) correction is suppressed by a factor of $\left(m/M\right)^n$ when compared to geodesic motion.

Developing a systematic framework for controlled calculations of $m/M$ corrections to binary dynamics is quite challenging. At 1SF, Refs.~\cite{Mino:1996nk,Quinn:1996am} obtained an equation of motion for a point particle interacting with its own gravitational field. The framework was further formalized in Refs.~\cite{Detweiler:2000gt,Detweiler:2002mi,Gralla:2008fg,Pound:2009sm} and extended to 2SF order in Refs.~\cite{Rosenthal:2006iy,Detweiler:2011tt,Pound:2012nt,Gralla:2012db} (see also \cite{Barack:2018yvs,Pound:2021qin,Galley:2006gs,Galley:2008ih}). Corrections at 2SF order will be relevant for LISA~\cite{Hinderer:2008dm,Isoyama:2012bx,Burko:2013cca}. Current state-of-the-art computations include observables for eccentric orbits at 1SF~\cite{Barack:2010tm,Barack:2011ed} and waveforms for quasicircilar orbits at 2SF~\cite{Wardell:2021fyy}. There is also recent interest in studying SF corrections for scattering trajectories~\cite{Gralla:2021qaf,Barack:2022pde,Barack:2023oqp}. 
The worldline-based approach of Ref.~\cite{Galley:2008ih} extended the pioneering work of Goldberger and Rothstein~\cite{Goldberger:2004jt} to curved space, and the framework we develop here shares some of its features. 
However, our framework is based on QFT, which has its own merits and challenges. For example, one key difference is the focus on calculating scattering amplitudes, as opposed to deriving corrections to equations of motion.

There exists many interesting connections between exact curved spacetimes and amplitudes around flat space. 
Indeed, by resumming Feynman diagrams~\cite{Duff:1973zz} or using on-shell methods~\cite{Neill:2013wsa}, one may relate an appropriate scattering amplitude directly to the \sch metric. 
Refs.~\cite{Mougiakakos:2020laz,Jakobsen:2020ksu} and \cite{DOnofrio:2022cvn} followed an analogous approach for the Schwarzschild-Tangherlini and Reissner-Nordström-Tangherlini metrics respectively.
Similarly, there is growing evidence that amplitudes of spinning particles capture observables related to the Kerr black hole~\cite{Siemonsen:2019dsu,Guevara:2019fsj,Guevara:2018wpp,Chung:2018kqs,Chung:2019duq,Chung:2020rrz,Chen:2021kxt,Arkani-Hamed:2019ymq,Bern:2020buy,Kosmopoulos:2021zoq,Bern:2022kto,Aoude:2020onz,Aoude:2022trd,Aoude:2023vdk,Aoude:2022thd,Maybee:2019jus,Bjerrum-Bohr:2023jau}. 
Moreoever, amplitudes containing all-orders-in-$G$ information have been extracted from calculations in curved space~\cite{Adamo:2020qru,Cristofoli:2020hnk,Adamo:2021rfq,Adamo:2022mev,Adamo:2022rmp,Adamo:2022qci,Adamo:2023cfp}.

In this paper, we combine elements of EFT and scattering amplitudes in curved space to develop a framework for computing gravitational self-force corrections to classical binary dynamics. 
By expanding around the black-hole geometry sourced by the heavy particle, our interaction vertices contain information to all orders in $G$, and observables are organized as an expansion in powers of $m/M$.
The dynamics of the probe particle are derived from the classical limit of self-energy diagrams of a massive scalar (see \figref{fig:SFCounting}), which may be thought of as a manifestation of the fact that self-force corrections are due to the particle interacting with its own gravitational field.

An important aspect of our framework is the backreaction of the background spacetime, which we capture via light degrees of freedom that follow from symmetry breaking considerations~\cite{Low:2001bw,Watanabe:2013iia,Delacretaz:2014oxa}.
In particular, the spontaneous breaking of spacetime symmetries due to the black-hole geometry dictates the existence of Goldstone bosons, which become relevant beyond the geodesic limit. This line of analysis is based solely on aspects of symmetry in QFT and leads us to incorporate effective degrees of freedom that are consistent with the worldline approach~\cite{Goldberger:2004jt}.

Our framework is an analytic perturbative scheme designed to leverage the all-orders-in-$G$ information encoded in the background geometry, while still taking advantage of the efficient tools developed for flat-space calculations. Scattering amplitudes in curved space encode binary dynamics of the inspiral phase. They do not capture non-perturbative effects; rather, they are constructed to match the corresponding amplitudes in flat space when expanded in $G$.
While the focus of our work is not on reformulating the PM expansion, it is worth noting that our curved-space framework offers several advantages for computing amplitudes in flat space. 
For example, particular families of diagrams in flat space are automatically combined in curved space, yielding compact expressions.
Moreover, higher-order loop diagrams in flat space are mapped to lower-order loop diagrams in curved space which in some cases lead to fewer integrations (see, e.g., \figref{fig:Hourglass}). 
In fact, our formalism yields certain classes of Feynman graphs to all orders in $G$.

We perform a number of nontrivial checks. In particular, we calculate the 0SF two-point amplitude, which corresponds to the flat-space two-to-two amplitude, to all orders in $G$, finding agreement with Ref.~\cite{Cheung:2020gbf}. 
Furthermore, since our curved-space amplitudes match flat-space amplitudes at fixed order in $G$, we recompute the $\mathcal{O}\big(G^2\big)$ two-point amplitude and the integrand for the $\mathcal{O}\big(G^3\big)$ two-point amplitude, finding agreement with Refs.~\cite{Bjerrum-Bohr:2002gqz,Bern:2019nnu}. As building blocks to the above amplitudes we obtain the Compton amplitude and the single-graviton-emission amplitude, which also match the literature~\cite{Luna:2017dtq}.

The remainder of this paper is organized as follows: In \secref{sec:Preliminary} we give a summary of our formalism, and review some background material and conventions. In \secref{sec:0SF} we use our framework to study the 0SF dynamics of a probe particle around a \sch black hole. 
In \secref{sec:1SF} we introduce the backreaction of the background and obtain the Feynman rules relevant for 1SF observables, we discuss the general powercounting and we provide consistency checks. 
Finally, in \secref{sec:Conclusions} we conclude and present possible future directions.
\smallskip

\textbf{Note added}: While this paper was at a late stage we learned of the upcoming work~\cite{Cheung:2023lnj} on an effective field theory for extreme mass ratios using a curved-space worldline formalism. We thank the authors for coordinating the appearance of the manuscripts.

\section{Preliminary Remarks}
\label{sec:Preliminary}

In this section we give an overview of our setup, discuss a few aspects of QFT in curved space, and establish our conventions. 

\subsection{Summary of the Setup}
The standard approach for describing classical two-body dynamics using scattering amplitudes is to consider two massive scalar fields interacting through gravity. Here, we instead model the binary system by considering only a single scalar field but in a curved background spacetime. This captures the dynamics of a light probe particle moving around a heavy gravitational source. Moreover, we model the backreaction or recoil of the heavy source against the probe particle by including additional light degrees of freedom. 
These three key elements are described by the action
\begin{equation}
    S = \actionG + \actionL + \actionH\,,
    \label{eq:ActionTotal}
\end{equation}
where the subscripts correspond to `gravitational,' `light' and `heavy.' 

For the gravitational part of the action we take
\begin{align}
    \actionG &= S_\text{EH} + S_\text{GF}\,,
\end{align}
where $S_\text{EH}$ is the Einstein Hilbert action and $S_\text{GF}$ is a gauge-fixing term; for the explicit form of these terms see \eqtref{eq:GravActionExplicit}. $\actionG$ describes the dynamics of the graviton $h_{\mu\nu}$. In particular, we define the graviton by expanding the metric as
\begin{align}
\label{eq:GravitonDef}
    g_{\mu\nu} = \gb_{\mu\nu} + \kappa\, h_{\mu\nu} \,,
\end{align}
for some background metric $\gb_{\mu\nu}$, which we later specify to be the \sch metric.

For the probe particle we consider a real scalar field $\phi$ minimally coupled to gravity,
\begin{equation}
    \actionL = \int d^4x \sqrt{-g} \left( \frac{1}{2}
        g^{\mu\nu} \partial_\mu\phi \partial_\nu\phi - \frac{1}{2} m^2 \phi^2 \right) \,.
\end{equation}

Finally, $\actionH$ captures the dynamics of the light degrees of freedom associated with the backreaction of the heavy source. As we discuss in more detail in \secref{sec:LightDoF}, choosing a particular background fixes these light degrees of freedom, and $\actionH$ is the most general effective theory that describes their interactions. Restricting to the case of a \sch black hole, for the purposes of this paper it is sufficient to consider
\begin{align}
    \actionH = -M \int d\tau \,,
    \label{eq:ActionHeavy}
\end{align}
where $M$ is the mass of the black hole appearing in the metric (see \eqsref{eq:bgMetricExplicit}{eq:ABDef}) and $\tau$ is its proper time. The heavy-particle contribution from this action sources the background geometry $\gb$ through Einstein's equations. The geodesic motion of the probe is then seen as a consequence of the curved metric rather than the exchange of gravitons between the two particles.  
The remaining contributions of this action, which only contribute beyond 0SF, capture the deflection of the source.

\subsection{Quantum Field Theory in Curved Spacetime}

Let us briefly comment on a few salient features of QFT in curved spacetime that are relevant for our setup. For a comprehensive treatment we refer the reader to textbooks, e.g.~\cite{Birrell:1982ix,Parker:2009uva}.

Firstly, we restrict our analysis to stationary spacetimes, where we are able to define a Fock space similar to the one we are familiar with from flat space. In particular, given that the spacetime is static, there exists a time-like Killing vector  $\partial_t$ associated with time translations, and we can choose both our in and out Fock spaces to comprise of eigenstates of this Killing vector, i.e. states of definite energy. This means that the in and out Fock spaces share the same vacuum and that there is no spontaneous particle production. On the other hand, if the time-like vector that defines energy eigenstates in the far past was different from the one in the far future, then a single-particle in-state might have non-trivial overlap with a multiparticle out-state.

Secondly, we restrict our analysis to spacetimes that are asymptotically flat. The black-hole geometries that belong in this category are of particular interest for modeling two-body dynamics, and we focus here on the simplest case of a \sch black hole. By constructing the Fock spaces as described above, we ensure that the vacuum state reduces to the Minkowski one away from the black hole. Thus the perturbative expansion of our amplitudes matches the corresponding one in flat space.

Finally, we do not consider horizon phenomena. Black-hole geometries generically have horizons, and one might be able to extend a geometry accurate outside the horizon to one that is also valid past the horizon. Defining a vacuum state on the extended geometry results in a state that looks mixed to an asymptotic observer, given that they do not have access to the part of the state that describes the interior of the horizon. In particular, this state appears thermal such that the density matrix corresponds to a thermal average at the Hawking temperature~\cite{Wald:1975kc,Parker:1975jm}. Particle detectors placed away from the black hole would register particles in this state, a phenomenon known as Hawking radiation~\cite{Hawking:1975vcx}. 

In our analysis we are effectively removing the physics of the horizon by our choice of vacuum state, and by treating the black-hole geometry as a (resummed) perturbative expansion about flat space. 
It would be interesting to return to this point and extend our formalism to retain these effects. Specifically, retaining the physics of the horizon would potentially give access to non-perturbative phenomena of interest for the binary problem, such as the excitation of quasi-normal modes.

\subsection{Conventions}

In this subsection we collect the conventions used in this paper.
We adopt $\kappa^2 = 32 \pi G$. 
For the Riemann tensor we use
\begin{align}
    g^{\mu\nu}R_{\alpha\beta\gamma\nu} = \Gamma^{\,\mu}_{\,\,\,\alpha\gamma,\beta} - \Gamma^{\,\mu}_{\,\,\,\beta\gamma,\alpha} + \Gamma^{\,\lambda}_{\,\,\,\alpha\gamma} \Gamma^{\,\mu}_{\,\,\,\beta\lambda} - \Gamma^{\,\lambda}_{\,\,\,\beta\gamma} \Gamma^{\,\mu}_{\,\,\,\alpha\lambda} \,.
\end{align}
For our scattering amplitudes we use the all-incoming convention. Throughout the paper we refer to the $n$SF two-point amplitude in curved space, which corresponds to the $n$SF two-to-two amplitude in flat space, simply as the $n$SF amplitude.
We work with a mostly-minus metric,
\begin{align}
    \eta_{\mu\nu} = 
    \begin{pmatrix}
        1 & 0 & 0 & 0\\
        0 & -1 & 0 & 0\\
        0 & 0 & -1 & 0\\
        0 & 0 & 0 & -1
    \end{pmatrix}\,,
    \qquad
    \eta^{\mu\nu} = 
    \begin{pmatrix}
        1 & 0 & 0 & 0\\
        0 & -1 & 0 & 0\\
        0 & 0 & -1 & 0\\
        0 & 0 & 0 & -1
    \end{pmatrix} \,.
\end{align}
We express the \sch metric in isotropic coordinates as follows:
\begin{align}
\label{eq:bgMetricExplicit}
    \gb_{\mu\nu} = 
    \begin{pmatrix}
        \gA & 0 & 0 & 0\\
        0 & -\gB & 0 & 0\\
        0 & 0 & -\gB & 0\\
        0 & 0 & 0 & -\gB
    \end{pmatrix}\,,
    \qquad
    \gb^{\mu\nu} = 
    \begin{pmatrix}
        \frac{1}{\gA} & 0 & 0 & 0\\
        0 & -\frac{1}{\gB} & 0 & 0\\
        0 & 0 & -\frac{1}{\gB} & 0\\
        0 & 0 & 0 & -\frac{1}{\gB}
    \end{pmatrix} \,,
\end{align}
where 
\begin{align}
\label{eq:ABDef}
    \gA = \frac{\left(1-a_4\right)^2}{\left(1+a_4\right)^2} \,,
    \quad
    \gB = \left(1+a_4\right)^{b_4} \,,
    \quad
    a_4 = \frac{G M}{2 r} \,,
    \quad
    b_4 = 4 \,.
\end{align}
The exposition given is in exactly four dimensions. When we need to regulate infinite intermediate expressions, we choose dimensional regularization where all expressions are lifted to $d$ dimensions. In particular, the black-hole geometry is also lifted to a corresponding $d$-dimensional version~\cite{Tangherlini:1963bw}. In this case $\eta_{\mu\nu}$ and $\gb_{\mu\nu}$ are diagonal $(d \times d)$ matrices of the above form, with 
\begin{align}
\label{eq:DimRegDef}
    a_4 \rightarrow a_d = \frac{4\pi G M}{(d-2) \Omega_{d-2}}\frac{1}{r^{d-3}} \,, 
    \quad
    b_4 \rightarrow b_d = \frac{4}{d-3} \,,
    \quad
    \Omega_{d-2} = \frac{2 \pi^{(d-1)/2}}{\Gamma\left(\frac{d-1}{2}\right)} \,.
\end{align}
%

\section{0SF: Geodesic Dynamics}
\label{sec:0SF}

In this section we describe the dynamics of the probe particle to all orders in $G$ in the limit where the source is infinitely heavy, i.e., $m \ll M$. This limit is referred to as 0SF order, and describes geodesic motion where the black hole is static during the scattering process. 

In the SF expansion, higher-SF orders are suppressed compared to lower-SF orders by factors of the small ratio $m/M$. In \secref{sec:FeynRulesNSF} we give the precise powercounting rules for our formalism, which make the SF counting apparent. For now we simply state that to describe geodesic motion we may neglect the dynamical graviton as well as the backreaction of the background, and  
therefore, 0SF dynamics is captured by 
\begin{align}
\label{eq:ActionLight0}
    \actionL^{(0)} \equiv \actionL\rvert_{g \rightarrow \gb} = \int d^4x \sqrt{-\gb} \left( \frac{1}{2}
        \gb^{\mu\nu} \partial_\mu\phi \partial_\nu\phi - \frac{1}{2} m^2 \phi^2 \right) \,,
\end{align}
where $\actionL = \sum_n \actionL^{(n)}$, with $\actionL^{(n)}$ being exactly order $n$ in the graviton field $h_{\mu\nu}$.
%

\subsection{0SF Amplitude at $\mathcal{O}(G)$}
\label{sec:Amplitude0SFLinearG}

Before tackling 0SF to all orders in $G$, we first demonstrate our formalism by deriving the two-point amplitude at $\mathcal{O}(G)$. 
We perform all steps of the calculation explicitly, and discuss the connection to the usual flat-space four-point amplitude. 

In order to use standard methods for calculating amplitudes, 
we need the propagators of our fields to be identical to those in flat space. We therefore identify 
\begin{align}
\label{eq:Lagrangian0SFSplit}
    \mathcal{L}_\text{L, free}^{(0)} &= 
        \frac{1}{2} \eta^{\mu\nu} \partial_\mu\phi \partial_\nu\phi - \frac{1}{2} m^2 \phi^2 \,, \nn \\
    \mathcal{L}_\text{L, int}^{(0)} &= \frac{1}{2} \left( \sqrt{-\gb} \gb^{\mu\nu} - \eta^{\mu\nu} \right) \partial_\mu \phi \partial_\nu \phi 
        -\frac{1}{2} m^2\left( \sqrt{-\gb}-1 \right) \phi^2 \,,
\end{align}
where $\actionL^{(0)}=\int d^4x \left( \mathcal{L}_\text{L,free}^{(0)} + \mathcal{L}_\text{L,int}^{(0)} \right)$. 
The terms appearing in $\mathcal{L}_\text{L, int}^{(0)}$ vanish in the $G\rightarrow 0$ limit as well as in the $r\rightarrow \infty$ limit, which justifies identifying them as interaction vertices. On the other hand, $\mathcal{L}_\text{L, free}^{(0)}$ is the usual free flat-space Lagrangian for a scalar particle, which implies the propagator
\begin{align}
  \scalarPropagator = \frac{\imUnit}{p^2 - m^2}\,.
\end{align}
Throughout this paper we have the $(\imUnit \epsilon)$ prescription implicit. Since we are calculating amplitudes, we choose the Feynman prescription. 
By splitting the Lagrangian in this manner, we have the same propagator as in flat space, and the all-orders-in-$G$ information is encoded in the interaction vertices. A similar approach was taken in Ref.~\cite{Komissarov:2022gax} in the cosmological context.

\begin{figure}
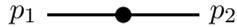

    \centering
    \scalarBgVertex
    \caption{Feynman graph for the probe two-point vertex obtained from $\actionLInt^{(0)}$ or $\actionLYInt^{(0)}$. We depict the probe particle as a black solid line. The dot signifies the background insertion. The background injects spatial three-momentum through the vertex, such that $\bm p_1 + \bm p_2 \neq 0$.}
    \label{fig:scalarBgVertex}
\end{figure}

In our setup, the existence of the heavy particle is captured by the interaction vertices. In particular, there is no field that creates a heavy-particle state, and thus there are no external states for the heavy particle in our amplitudes. We therefore have the heuristic map $\mathcal{A}^\text{curved}_n \sim \mathcal{A}^\text{flat}_{n+2}$ between $n$-point amplitudes in curved space and $(n+2)$-point amplitudes in flat space. An important part of this work focuses on exploring the details of this map.
 
For two-body dynamics, we are interested in the two-point amplitude for a probe particle moving in the background metric $\gb$, which corresponds to the four-point flat-space amplitude between the probe and the heavy particle that sources $\gb$. We obtain the two-point amplitude using the Feynman rule derived from $\mathcal{L}_\text{L, int}^{(0)}$ and shown in \figref{fig:scalarBgVertex}. We define
\begin{align}
    C_\text{L}^{(0)\mu\nu} = \frac{1}{2} \left( \sqrt{-\gb} \gb^{\mu\nu} - \eta^{\mu\nu} \right)\,,
    \quad 
    C_\text{L}^{(0)} = -\frac{m^2}{2} \left( \sqrt{-\gb}-1 \right) \,,
\end{align}
such that,
\begin{align}
    \mathcal{L}_\text{L, int}^{(0)} = C_\text{L}^{(0)\mu\nu} \partial_\mu \phi \partial_\nu \phi 
        + C_\text{L}^{(0)} \phi^2 \,.
\end{align}
In momentum space, we have 
\begin{align}
    \actionLInt^{(0)} = \int d^4 x \mathcal{L}_\text{L, int}^{(0)}  
    = \int \frac{d^4 p}{(2\pi)^4} &\frac{d^4 p^\prime}{(2\pi)^4}\frac{d^3 \bm q}{(2\pi)^3} (2\pi) \delta\left( E +E^{\prime} \right) (2\pi)^3 \delta^{(3)}\left( \bm p + \bm p^\prime + \bm q \right) \times \nn \\
    & \tilde{\phi}(p) \tilde{\phi}(p^\prime) \left( -p_{\mu} p^\prime_{\nu} \tilde{C}_\text{L}^{(0)\mu\nu}(\bm q) + \tilde{C}_\text{L}^{(0)}(\bm q) \right)  \,,
\end{align}
where $p = (E,\bm p)$, $p^\prime = (E^\prime,\bm p^\prime)$, and we denote functions related by a Fourier transform with a tilde.
The two-point vertex respects energy conservation for the two incoming scalar lines, but spatial three-momentum is not conserved; the background inserts spatial momentum through the vertex. This agrees with the fact that the background metric respects time-translation symmetry but breaks spatial-translation symmetry (see also \secref{sec:LightDoF}). 

We may now straightforwardly compute the two-point amplitude with a single insertion of the interaction vertex in \figref{fig:scalarBgVertex}. We find
\begin{align}
    \mathcal{A}_2^\text{s.in.}(p_2,\bm q) = 
        2 p_{2\mu} (p_{2\nu}+q_\nu) \tilde{C}_\text{L}^{(0)\mu\nu}(\bm q) + 2 \tilde{C}_\text{L}^{(0)}(\bm q) \,,
\end{align}
where $q=(0,\bm q)$ is the momentum transfer and `s.in.' stands for `single insertion.' As in flat space we strip off a factor of $ 2\pi  \imUnit \,  \delta\left( E_2 + E_3 \right) $ in defining the amplitude, noting that here we do not have three-momentum conservation and hence only have a delta function for energy conservation.
While we cannot obtain $\tilde{C}_\text{L}^{(0)\mu\nu}$ and $\tilde{C}_\text{L}^{(0)}$ exactly, we may perform the necessary Fourier transform order by order in $G$.
In particular, to leading order in $G$ and in the classical limit we have
\begin{align}
\label{eq:Leading0SFAmp}
    \mathcal{A}_2^\text{s.in.}(p_2,\bm q) = \frac{8 \pi G M m^2  \left(1-2 \gamma^2\right)}{\bm q^2} + \mathcal{O}\left(G^2\right)  \,, 
\end{align}
where $\gamma = E_2/m$. This result matches the classical flat-space $2\rightarrow2$ amplitude at $\mathcal{O}\left(G\right)$ for two scalar particles interacting via graviton exchange, evaluated in the appropriate frame (we discuss this frame choice in detail in \secref{sec:OneSFAmplitude}).
Note that in order to match the flat-space amplitude with relativistic normalization for all particles, we need to multiply the curved-space amplitude with a factor of $2M$ for each background-insertion vertex.

\subsection{0SF Amplitude to All Orders in $G$}
\label{sec:Amplitude0SFAllG}

We now present a novel derivation of the 0SF contribution to all orders in $G$ by directly computing amplitudes in curved space. It is straightforward to extend the calculation described in the previous section to compute diagrams with $n$ insertions of the two-point vertex, and then resum the infinite set of contributions as a geometric series. Here, we present an alternative approach where the complete 0SF amplitude is instead given in terms of a single-insertion diagram using a different field basis.

The idea is to perform a field redefinition such that any multi-insertion diagram gives only superclassical iteration contributions. We start from $\actionL^{(0)}$ given in Eq.~(\ref{eq:ActionLight0}) and plug in the explicit form of the background metric given in \eqtref{eq:bgMetricExplicit},
\begin{align}
\actionL^{(0)} = \int d^4 x \frac{\gB \sqrt{\gA \gB}}{2} \left(
	\frac{\left( \partial_t \phi \right)^2}{\gA} - \frac{(\nabla \phi)^2}{\gB} -
	m^2 \phi^2 \right) \,.
\end{align}
We then perform the field redefinition
\begin{equation}
\label{eq:FieldRedY}
\phi \rightarrow \frac{\phi}{\left( \gA \gB \right)^{1/4}}\,,
\end{equation}
which results in
\begin{align}
\label{eq:ActionLightY0Def}
\actionL^{(0)} \rightarrow \actionLY^{(0)} &= \int d^4 x \, \frac{1}{2} \Bigg(
	\frac{\gB}{\gA} \left( \partial_t \phi \right)^2 
	-(\nabla \phi)^2 
	+ \left( -\gB m^2 + F \right) \phi^2
		\Bigg) \,, \nn \\
		F&=-\frac{\nabla^2 \gA}{4 \gA}+\frac{3 (\nabla\gA)^2}{16 \gA^2}
		-\frac{(\nabla\gA) \cdot (\nabla\gB)}{8 \gA \gB}
		+\frac{3 (\nabla\gB)^2}{16 \gB^2}-\frac{\nabla^2\gB}{4 \gB} \,,
\end{align}
up to total derivatives. As we see below this new action naturally yields the amplitude in the so-called Y-pole subtraction scheme~\cite{Cheung:2018wkq}, hence the subscript label `LY.' Note that the field redefinition reduces to the identity in both the $G\rightarrow 0$ and the $r\rightarrow \infty$ limits, which means that it is a valid perturbative redefinition that does not change the normalization of the external states. We find that $F$ only contributes terms with subleading scaling in the classical limit, and we drop it henceforth.

Next, we split $\actionLY^{(0)}$ into a free part and an interaction part as in \eqtref{eq:Lagrangian0SFSplit}. The free part is the same as before. The interaction part is
\begin{align}
\actionLYInt^{(0)} = \int d^4 x \, \frac{1}{2} \Bigg(
	\left( \frac{\gB}{\gA}-1\right) \left( \partial_t \phi \right)^2
	+ \left( -\gB m^2 + m^2  \right) \phi^2
		\Bigg) \,,
\end{align}
which again results in a Feynman vertex as the one depicted in \figref{fig:scalarBgVertex}.
As we noted, this two-point vertex does not impart energy to the probe particle. We may manifest this fact by Fourier transforming in time,
\begin{align}
\actionLYInt^{(0)} = \int d^3 \bm x \frac{d E}{2\pi} \frac{d E^\prime}{2\pi}  &(2\pi) \delta(E+E^\prime) \phi(E,\bm x) \phi(E^\prime, \bm x) 
	\times \nn \\
	& \, \frac{1}{2} \Bigg(
	E^2 \left( \frac{\gB}{\gA}-1\right) + \left( -\gB m^2 + m^2 \right)
		\Bigg) \,.
\end{align}
It is straightforward to see that multiple insertions of this vertex only contribute to superclassical iterations: Such contributions involve propagators of the probe particle that need to be cancelled in order to yield a classical contribution. However, the field redefinition was chosen such that the vertex is 
free of spatial derivatives, and thus it is impossible to get the inverse propagators required for cancellation. 

We thus identify
\begin{equation}
\label{eq:Amp0SFPositionSpace}
\tilde{\mathcal{A}}_{2}^\text{0SF}(p_2, \bm x) = m^2 \left( \gamma^2 \left( \frac{\gB}{\gA}-1\right) + \left( 1-\gB \right) \right) \,,
\end{equation}
as an off-shell position-space generating function for the complete 0SF amplitude. The 0SF amplitude in momentum space is then
\begin{align}
    \mathcal{A}_{2}^\text{0SF}(p_2,\bm q) = \left( \int \frac{d^3 \bm x}{(2\pi)^3} e^{-\imUnit \bm q \cdot \bm x} \tilde{\mathcal{A}}_{2}^\text{0SF}(p_2, \bm x) \right)_\text{on shell} + \text{iteration} \,,
\end{align}
where by `on shell' we mean that $\bm q$ needs to be restricted to obey $\bm p_2 \cdot \bm q = -\bm q^2/2$ and `iteration' denotes the superclassical iteration contributions organized in terms of Y poles~\cite{Cheung:2018wkq}. The above generating function was obtained in Ref.~\cite{Cheung:2020gbf}, and an 
equivalent result was obtained in Ref.~\cite{Kalin:2019rwq}.

We may readily expand $\tilde{\mathcal{A}}_2^\text{0SF}(p_2, \bm x)$ to any order in $G$ and perform the Fourier transform to obtain the classical part of $\mathcal{A}_{2}^\text{0SF}(p_2,\bm q)$ to that order. Already in position space, however, we recognize the familiar combinations of $\gamma^2$ that appear in the probe limit of the flat-space scattering amplitudes (see, e.g., Eq.~(9.3) of Ref.~\cite{Bern:2019crd}). To the first few orders we have,
\begin{align}
\label{eq:PositionSpaceAmpExplicit}
    \tilde{\mathcal{A}}_{2}^\text{0SF}(p_2, \bm x) &= 
    \frac{\left(2 \gamma ^2-1\right) G M m^2 }{r}
    +\frac{3 \left(5 \gamma ^2-1\right) G^2 M^2 m^2 }{4 r^2} \nn \\
    &+\frac{\left(18 \gamma ^2-1\right) G^3 M^3 m^2 }{4
   r^3}
   +\frac{\left(129 \gamma ^2-1\right) G^4 M^4 m^2 }{32 r^4}
   + \mathcal{O}\left(G^5\right) \,.
\end{align}
%

\subsection{Heavy Expansion and Geodesic Expansion}

We conclude this section with a brief discussion of two alternative setups for analyzing the probe particle. The first setup is the heavy expansion familiar from Heavy Quark Effective Theory~\cite{Wise:1992hn} as well as its recent applications for the binary problem~\cite{Damgaard:2019lfh}. The heavy particle expansion for the real scalar field~\cite{Heinonen:2012km} representing the probe particle is 
\begin{align}
    \phi(x) \, = \, e^{-\imUnit p \cdot x} \varphi_p(x)/\sqrt{2m}
    \, = \, e^{\imUnit p \cdot x} \varphib_p(x)/\sqrt{2m} \,.
\end{align}
With this expansion, every term in the Lagrangian has definite powercounting in $m$ since derivatives on $\varphi_p(x)$ and $\varphib_p(x)$ are $\mathcal{O}\left(q\right)$. We expect that with an appropriate field redefinition, this heavy expansion leads to amplitudes that are directly expressed in terms of Z-poles~\cite{Parra-Martinez:2020dzs,Bern:2021dqo} instead of Y-poles.

The second setup is inspired by the first and is similar to the WKB approximation: instead of expanding around a heavy particle moving in a straight line, we expand around a heavy particle following a geodesic in curved space~\cite{Balasubramanian:2019stt} (see also Refs.~\cite{Kol:2021jjc,Adamo:2023cfp,Kim:2023qbl}). 
We thus consider the following ansatz where the plane-wave factor is replaced with a semi-classical solution to the equations of motion:
\begin{align}\label{eq:geodesicEXP}
    \phi(x) \, = \, e^{-\imUnit S_p(x)} \Phi_p(x)/\sqrt{2m} \, = \, e^{\imUnit S_p(x)} \bar{\Phi}_p(x)/\sqrt{2m} \,.
\end{align}
Here we take $S_p(x)$ to be the classical on-shell action satisfying the Hamilton-Jacobi equation, along with the requirement that it reduces to a plane wave in the limit of vanishing $G$,
\begin{align}
    \gb^{\mu\nu} \, \partial_\mu S_p(x) \, \partial_\nu S_p(x) = m^2
    \qquad \text{and} \qquad S_p(x) \rightarrow p \cdot x \quad \text{as} \quad G \rightarrow 0 \,.
\end{align}
We may iteratively compute $S_p(x)$ in terms of simple one-dimensional integrals. The explicit form to the first few orders in $G$ is 
\begin{align}\label{eq:ActionWKB}
    S_p(x) &= m \left(\gamma \,  t+\sqrt{\gamma ^2-1} \, z\right)
    + \frac{m G M \left(2 \gamma ^2-1\right) \tanh ^{-1}\left(\frac{z}{r}\right)}{\sqrt{\gamma ^2-1}} \\
    &-\frac{m G^2 M^2 \left(2 \left(1-2 \gamma ^2\right)^2 z-3 \left(5 \gamma^4 -6 \gamma^2+1\right) \sqrt{r^2-z^2} \tan^{-1}\left(\frac{z}{\sqrt{r^2-z^2}}\right)\right)}{4 \left(\gamma^2-1\right)^{3/2} \left(r^2-z^2\right)} \nn \\
    &+\mathcal{O}\left(G^3\right) \,, \nn 
\end{align}
where we took $p_\mu = m \left(\gamma,0,0,\sqrt{\gamma ^2-1}\right)$. The 0SF scattering angle follows from this on-shell action using standard methods of classical mechanics.

The expansion in Eq.~(\ref{eq:geodesicEXP}) may be useful for resumming geodesic motion and systematically calculating deviations from it. An encouraging observation is that using this expansion in the Lagrangian removes the leading-in-$m$ term, effectively factoring out the 0SF contribution. We postpone a detailed investigation of this to future work.

\section{Beyond 0SF: Backreaction}
\label{sec:1SF}

Beyond 0SF, the motion of the probe perturbs spacetime, and induces a backreaction in the background metric. This backreaction is mediated by the graviton, and also involves the deflection of the heavy source. In this section we analyze these key elements, and describe the Feynman and powercounting rules necessary for computing the $n$SF amplitude. We apply our formalism for computing the 1SF amplitude at ${\cal O}(G^2)$ and ${\cal O}(G^3)$, providing a non-trivial consistency check of the framework developed in this paper.

\subsection{Purely Gravitational Part of the Action}

The purely gravitational part of the action is given by the Einstein-Hilbert term and the gauge-fixing term:
\begin{alignat}{3}
\label{eq:GravActionExplicit}
    \actionG = S_\text{EH} + S_\text{GF} &=  
    -\frac{1}{16 \pi G} \int d^4x \sqrt{-g} &&R 
    &&+\int d^4x \sqrt{-\gb} F^\mu F_\mu \,,  \nn \\
    F_\mu &= \cov^\nu h_{\mu\nu}-\frac{1}{2} \cov_\mu h \,, 
    &&h &&= \gb^{\alpha\beta} h_{\alpha\beta} \,,
\end{alignat}
where $\cov_\mu$ is the background-metric covariant derivative, indices are raised and lowered with the background metric $\gb$, and the graviton field $h_{\mu\nu}$ is defined in~\eqtref{eq:GravitonDef}.  
We emphasize that the complete graviton dynamics in curved space includes its interaction with the light degrees of freedom of the heavy source, which we describe in~\secref{sec:LightDoF}. 

We obtain Feynman rules from $S_G$ by expanding the metric as in~\eqtref{eq:GravitonDef}. The resulting linear-order-in-$h$ term cancels against the corresponding term in $\actionH$, as we demonstrate in \secref{sec:LinearOrderFluctuations}. The higher-order-in-$h$ terms yield the propagator for $h$ as well as interaction vertices with the background, similar to our analysis in \secsref{sec:Amplitude0SFLinearG}{sec:Amplitude0SFAllG}. Note that we only need vertices with up to $(n+1)$ graviton fields for the calculation of the $n$SF amplitude. 
In particular, in this paper we explicitly consider dynamics up to 1SF order and hence only need the action to quadratic order in $h$:
\begin{align}
\label{eq:ActionGravityQuadratic}
    \frac{\mathcal{L}_\text{G}}{\sqrt{-\bar{g}}} &= 
    \frac{\kappa}{16 \pi G} \bar{G}^{\alpha\beta} h_{\alpha\beta}
    - \frac{1}{4} h^{ ; \beta}h_{;\beta} 
    + \frac{1}{2} h_{\alpha\beta ; \gamma}h^{\alpha\beta ; \gamma} 
    \\
	&+ \frac{1}{4} \bar{R} \left( 2 h^{\alpha\beta} h_{\alpha\beta} - h^2 \right) 
    + \bar{R}_{\beta\gamma} \left( h^{\beta\gamma} h
        - h_\alpha^\gamma h^{\alpha\beta}  \right) 
    - \bar{R}_{\alpha\gamma\beta\lambda} h^{\alpha\beta} h^{\gamma\lambda}
	+ \mathcal{O}\left(h^3\right)
    \,. \nn
\end{align}
Here $\actionG=\int d^4x \mathcal{L}_\text{G}$ and we use the shorthand $(\ldots)_{;\mu} \equiv \cov_\mu (\ldots)$ for the covariant derivative.
The Einstein tensor, Ricci scalar, Ricci tensor and Riemann tensor are evaluated on the background metric, and are respectively denoted by $\bar{G}^{\alpha\beta}$, $\bar{R}$, $\bar{R}_{\beta\gamma}$ and $\bar{R}_{\alpha\gamma\beta\lambda}$.

We do not set the Einstein tensor, Ricci scalar and Ricci tensor to zero, despite the fact that $\gb$ is eventually taken to be the \sch metric. 
For instance, for the linearized \sch metric, $\bar{G}^{\alpha\beta}$, $\bar{R}$ and $\bar{R}_{\beta\gamma}$ evaluate to expressions proportional to a delta function with support on the position of the black hole. 
These terms are on a similar footing as terms originating from $\actionH$, which also have support on the position of the black hole; see \secref{sec:LinearOrderFluctuations}.
%

\subsection{Light Degrees of Freedom of the Heavy Source}
\label{sec:LightDoF}

To capture the backreaction of the metric we need to introduce dynamics for the light degrees of freedom of the heavy source. While the heavy source could have various light degrees of freedom, the minimum ones are those dictated by symmetry considerations. 
For the \sch black-hole geometry in \eqtref{eq:bgMetricExplicit},
the existence of a source at rest in the vacuum state spontaneously breaks Poincar\'e symmetry to rotational symmetry. The breaking of the three translation generators induces three Goldstone bosons $\zeta^i(t)$, where $i=1,2,3$ 
and $t$ is the time coordinate in the system defined by \eqtref{eq:bgMetricExplicit}, while the breaking of the boost generators does not induce more Goldstone bosons~\cite{Low:2001bw,Watanabe:2013iia}. The $\zeta^i(t)$ describe the fluctuations of the heavy source away from its original position, hence we refer to them as deflections.

The most general effective theory that captures the dynamics of the $\zeta^i(t)$ follows from the coset construction~\cite{Delacretaz:2014oxa}. Here we only consider minimal coupling to gravity, which leads to \eqtref{eq:ActionHeavy}, but it is straightforward to include finite-size effects.
The resulting action is that of a point particle coupled to gravity, e.g., as in Ref.~\cite{Goldberger:2004jt}, which is intuitive given the interpretation of the $\zeta^i(t)$ as the displacement of the heavy source.
Following standard manipulations (see, e.g., Sec.~1.2 of~\cite{Polchinski:1998rq}), we embed $\zeta^i(t)$ into a covariant but redundant description 
\begin{align}
   \zeta^i(t) \rightarrow z^\mu(\tau) \,,
   \label{eq:redundancy}
\end{align}
in terms of which we may trade \eqtref{eq:ActionHeavy} for the einbein-gauge-fixed Polyakov form,
\begin{align}
\label{eq:PolyakovActionHeavy}
    \actionH = -\frac{M}{2} \int d\tau \left( g_{\mu\nu} \dot{x}^\mu \dot{x}^\nu + 1 \right)\,,
    \quad 
    x^\mu = v^\mu \tau + z^\mu \,,
\end{align}
which is more convenient for computations. We take $v^\mu = (1,0,0,0)$.
In the full quantum theory, $S_H$ also includes gauge-fixing terms related to the reparametrization invariance of the worldline, which itself is related to the redundancy introduced in \eqtref{eq:redundancy} (see, e.g., Sec.~4.2 of~\cite{Polchinski:1998rq}). These terms, however, are not needed for our analysis, which follows closely that of Ref.~\cite{Mogull:2020sak}. 

The action in Eq.~(\ref{eq:PolyakovActionHeavy}) is ill-defined when expanding around the gravitational field sourced by the heavy particle. Indeed, the background metric has a singularity at the position of the heavy particle, which is where we evaluate the action. Expanding \eqtref{eq:PolyakovActionHeavy} as in \eqtref{eq:GravitonDef} and dropping the constant term we have
\begin{align}
    \actionH = -\frac{M}{2} \int d\tau \left( \gb_{\mu\nu} \dot{x}^\mu \dot{x}^\nu + \kappa h_{\mu\nu} \dot{x}^\mu \dot{x}^\nu \right) \,,
\end{align}
where $\gb$ and $h$ are evaluated at $x$. 
We define $\delta \gb \equiv \gb - \eta$ and isolate the divergent term as follows:
\begin{align}
\label{eq:ActionHeavyRegular}
    \actionH &= S_\text{H, div} + \actionHFin \,, \nn \\
    S_\text{H, div} &= -\frac{M}{2} \int d\tau \, \delta\gb_{\mu\nu} \dot{x}^\mu \dot{x}^\nu \,, \nn \\
    \actionHFin &= -\frac{M}{2} \int d\tau \left( \eta_{\mu\nu} \dot{x}^\mu \dot{x}^\nu + \kappa h_{\mu\nu} \dot{x}^\mu \dot{x}^\nu \right) \,.
\end{align}
For our analysis, which focuses on classical contributions in the potential region, we only require the finite term $S_\text{H, fin}$.

Here we consider the special case of the \sch black hole, while it would be interesting to extend our analysis to the Kerr black hole. The degrees of freedom and corresponding action for a vacuum state that spontaneously breaks rotational symmetries along with translational ones have been categorized in Ref.~\cite{Delacretaz:2014oxa}. Alternatively, we could work with a worldline theory along the lines of Refs.~\cite{Porto:2005ac,Kalin:2020mvi,Liu:2021zxr}. We postpone this analysis to future work. 

\subsection{Stability of the Background}
\label{sec:LinearOrderFluctuations}

In this subsection we demonstrate that our background is a proper vacuum state. In other words, our field configuration solves the equations of motion, or, equivalently, the Lagrangian terms linear in the fluctuations $h$ and $z$ vanish. 

We start with the linear-in-$z$ terms, which arise from the first term of $\actionHFin$ in \eqtref{eq:ActionHeavyRegular}. We have
\begin{align}
    \actionHFin\big\vert_{\text{linear-in-}z} = -M \int d\tau \eta_{\mu\nu} v^\mu \dot{z}^\nu \, .  
\end{align}
This is a total derivative, and hence may be removed from the action.

Next, we consider the terms linear in the graviton field $h$. We have two contributions, one from $\actionG$ and one from $\actionHFin$:
\begin{align}
\label{eq:ActionEinsteinEquation}
    \actionG \big\vert_{\text{linear-in-}h} &= 
    \int d^4 x \sqrt{-\gb} \left( \frac{1}{16 \pi G} \bar{G}^{\mu\nu} \right) \kappa h_{\mu\nu} \,, \nn \\
    \actionHFin \big\vert_{\text{linear-in-}h} &= 
    \int d\tau \left( -\frac{1}{2} M v^\mu v^\nu \right) \kappa h_{\mu\nu}
    \,.
\end{align}
For the background configuration to be stable, we need these two contributions to cancel. Here we see that the Einstein tensor $\bar{G}^{\mu\nu}$ cannot be set to zero; instead, these equations hint that it should be thought of as a distribution with support at the origin. This is indeed sensible to linear order in $G$, while there are subtleties at higher orders~\cite{Geroch:1986jjl}. 

In practice, to establish that the two terms in \eqtref{eq:ActionEinsteinEquation} cancel at leading order in $G$, we need to compute tensors such as $\bar{G}^{\mu\nu}$ with support at the origin. We do this by taking their Fourier transform in dimensional regularization. For the present paper, using the linear-order-in-$G$ background metric in $\actionG$ is sufficient, and we consider the stability of the background to this order only. It would also be interesting to carry out the analysis beyond this order.

\subsection{Feynman Rules at 1SF}

We now proceed to obtain the Feynman rules relevant for the calculation of the 1SF amplitude.
The Feynman rules in curved space are not more complicated than the ones in flat space even though they encode information to all orders in $G$. 

We start by considering the graviton vertices originating in $\actionG$. Similarly to our treatment in \secref{sec:0SF}, we split the action given in \eqtref{eq:ActionGravityQuadratic} into a free and interacting part by isolating the flat-space limit. The free part is 
\begin{align}
	\actionGFree = \int d^4 x \left(
        \frac{1}{2} \eta^{\mu\nu}\eta^{\alpha\gamma}\eta^{\beta\delta} 
        - \frac{1}{4} \eta^{\mu\nu} \eta^{\alpha\beta} \eta^{\gamma\delta} 
        \right)
        \partial_\mu h_{\alpha\beta} \partial_\nu h_{\gamma\delta}  \,,
\end{align}
which yields the usual de Donder propagator for the graviton,
\begin{align}
	\gravitonPropagator &= \frac{\imUnit}{q^2} \mathcal{P}^{\mu\nu\alpha\beta} \,, \quad
	\mathcal{P}^{\mu\nu\alpha\beta} = \frac{1}{2} \left( \eta^{\mu\alpha}\eta^{\nu\beta} + \eta^{\nu\alpha}\eta^{\mu\beta} - \eta^{\mu\nu}\eta^{\alpha\beta} \right) \,.
\end{align}
As mentioned above, we keep the $(\imUnit \epsilon)$ prescription implicit. The interaction vertices are obtained from the remainder
\begin{align}
	\actionGInt = \actionG - \actionGFree \,.
\end{align}
We depict the Feynman graph for the two-point graviton vertex in \figref{fig:gravitonBgVertex}. 

\begin{figure}
    \centering
    \gravitonBgVertex
    \caption{Feynman graph for the graviton two-point vertex obtained from $\actionGInt$, where the gravitons are depicted as wiggly lines. The dot signifies the background insertion. The background injects spatial momentum through the vertex, such that $\bm q_1 + \bm q_2 \neq 0$.
    }
    \label{fig:gravitonBgVertex}
\end{figure}

\begin{figure}
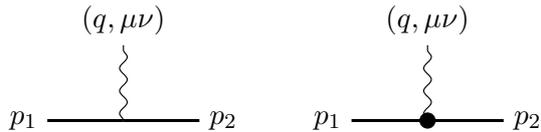

     \centering
     \scalarGravitonVertex
     \caption{Feynman graphs for the probe-graviton three-point vertices obtained from $\actionL^{(1)}$ or $\actionLY^{(1)}$. The left vertex is the same as one would obtain in flat space, hence the usual momentum conservation holds. The dot on the right one signifies the background insertion. The background injects spatial momentum through the vertex, such that $\bm p_1 + \bm p_2 + \bm q \neq 0$. The solid black lines stand for the probe particle, while the wiggly lines stand for the graviton.}
     \label{fig:ProbeGravitonVertex}
 \end{figure}

Next, we consider the coupling of the graviton to the probe particle.  We expand $\actionL$ as in \eqtref{eq:GravitonDef} and keep terms linear in $h_{\mu\nu}$:
\begin{align}
\label{eq:LightMatterActionFirstOrder}
   \actionL^{(1)} = \kappa \int d^4 x \sqrt{-\gb} \, h^{\mu\nu} \left(
   	- \frac{1}{2} \partial_\mu \phi \partial_\nu \phi
   	+ \frac{1}{4} \gb_{\mu\nu} \gb^{\alpha\beta} \partial_\alpha \phi \partial_\beta \phi
   	- \frac{1}{4} \gb_{\mu\nu} m^2 \phi^2
        \right) \,.
\end{align}
 Here we used the notation introduced below \eqtref{eq:ActionLight0}. Similar to \eqtref{eq:ActionLightY0Def}, we obtain $\actionLY^{(1)}$ via the field redefinition in \eqtref{eq:FieldRedY}, and then derive the probe-graviton interaction vertex, which we depict in \figref{fig:ProbeGravitonVertex}. For convenience, we split the vertex into the usual flat-space three-point vertex, and the remainder, which involves the background.

\begin{figure}
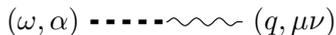

     \centering
     \deflectionGravitonVertex
     \caption{Feynman graph for the deflection-graviton two-point vertex obtained from $\actionHFinInt$. The dashed line stands for the deflection field, while the wiggly stands for the graviton. The vertex satisfies energy conservation: $\omega+q_0=0$.}
     \label{fig:DeflectionGravitonVertex}
 \end{figure}

Finally, we consider the deflection vertices, following the treatment in Ref.~\cite{Mogull:2020sak}. Again we split $\actionHFin$ into a free part $\actionHFinFree$ and an interacting part $\actionHFinInt$:
\begin{align}
	\actionHFinFree &= -\frac{M}{2} \int d\tau \, \eta_{\mu\nu} \dot{z}^\mu \dot{z}^\nu \,, \quad
	\actionHFinInt = \actionHFin - \actionHFinFree \,.
\end{align}
The free part gives the propagator of the deflection field,
\begin{align}
	\deflectionPropagator = -\frac{\imUnit}{M \omega^2}\eta^{\alpha\beta} \,,
\end{align}
where, again, based on the observable considered, one might need to keep track of the $(\imUnit \epsilon)$ prescription.
From the interacting part of the action we extract the vertex sufficient for computing the 1SF amplitude. In doing so, we keep in mind that the worldline Lagrangian is evaluated at $x^\mu$ given in \eqtref{eq:PolyakovActionHeavy}. We find
\begin{align}
\label{eq:actionHFinInt}
\actionHFinInt = -\frac{\kappa M}{2} \int d\tau \left( 
        2  h_{\mu\nu}(v\tau) v^\mu \dot{z}^\nu 
        + h_{\mu\nu,\alpha}(v\tau) v^\mu v^\nu z^\alpha +\mathcal{O}\left(z^2\right) \right) \,,
\end{align}
where the graviton field is evaluated at $v^\mu \tau$. This yields the deflection-graviton vertex depicted in \figref{fig:DeflectionGravitonVertex}. Since $\actionHFinInt$ is independent of the background metric, the vertex is identical to the one of Ref.~\cite{Mogull:2020sak}.

\subsection{Diagrams at $n$SF}
\label{sec:FeynRulesNSF}

In this subsection we discuss a few aspects of the diagrams contributing to the $n$SF amplitude, namely their powercounting, the required Feynman rules, and the contribution from the deflection of the heavy source.

\begin{figure}
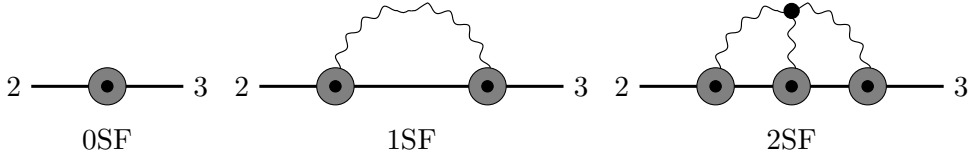

     \centering
     \SFCounting
     \caption{Schematic depiction of the diagram loop-order $\leftrightarrow$ SF-order correspondence in our formalism. Only representative diagrams are shown. 
     The gray blobs with the black dot denote the corresponding tree-level amplitudes where diagrams with any number of background insertions and virtual deflections have been included. 
     The external probe-particle lines have momenta $p_2$ and $p_3$.}
     \label{fig:SFCounting}
 \end{figure}

Firstly, we note that the diagram loop counting for the two-point amplitudes in curved space coincides with the SF counting, but it is different from the number of loop integrations as we discuss below. For example, for the 0SF dynamics considered in~\secref{sec:0SF} we only have diagrams with no loops, while the 1SF amplitudes considered in this section are given by diagrams with a single loop. This is schematically shown in~\figref{fig:SFCounting}. A similar relation between the loop and SF counting was noted in Ref.~\cite{Galley:2008ih}.

To see this, recall that in the classical limit the two-point amplitude $\mathcal{A}_2$ admits the expansion,
\begin{alignat}{2}
    &\text{Momentum space:} \qquad
    &&\mathcal{A}_2(\bm q) = \frac{G m^2}{\bm q^2} \sum_{n=1}^\infty 
        P_n\left(m,\,M\right)  \left( G |\bm q| \right)^{n-1} \,, 
    \nn \\
    &\text{Position space:} 
    &&\tilde{\mathcal{A}}_2(r) = m^2 \sum_{n=1}^\infty 
        \tilde{P}_n\left(m,\,M\right)
        \left( \frac{G}{r} \right)^n \,,
\end{alignat}
where $P_n$ and $\tilde{P}_n$ are polynomials of degree $n$ in the arguments and we keep the dependence of $p_2$ and $\gamma$ implicit. Neglecting the deflection contribution momentarily, factors of $M$ only appear through the background metric and only in the combination $G M/r$, which has classical scaling. Hence, in our formalism the amplitude is organized as
\begin{align}
    \tilde{\mathcal{A}}_2(r) &= m^2 \sum_{n=0}^\infty 
        \mathcal{F}_n\big[ G M / r \big]
        \left( \frac{G m}{r} \right)^n \,,
\end{align}
in position space and similarly in momentum space, for some functions $\mathcal{F}_n\big[ G M/r \big]$ (see, e.g., \eqtref{eq:PositionSpaceAmpExplicit}). In this way, we see that powers of $m$ are correlated to powers of $G$ that do not originate in background insertions. In fact, the source of these factors is the probe-graviton vertices, establishing that the effective coupling constant in our theory is $G m$, from which it immediately follows that loops count the SF order, i.e. the order in $m$. Finally, returning to the deflection contributions, these also scale classically and contribute via the combination $G M$, as can be seen, e.g., in \eqtref{eq:ComptonInParts}, so they are on the same footing as background insertions. 

We evaluate diagrams using the familiar flat-space tools, such that a diagram with $l$ loops, $n$ background insertions and $k$ virtual deflections requires $l + n + k - 1$ integrals. We emphasize again that the number of loops in the diagram and the number of loop integrals, while related, do not coincide in our formalism.

Secondly, we note that we only require a finite number of Feynman rules to compute diagrams at a given SF order since vertices with higher multiplicity would lead to more loops and would thus be higher SF order. In particular, at $n$SF, we require vertices with up to $(n+1)$ gravitons, vertices with two probe particles and up to $n$ gravitons, and vertices with a single graviton and $n$ deflections. In the previous subsection, we obtained the Feynman vertices sufficient for computing the 1SF amplitude.

Finally, for an analysis that focuses on the potential region for the gravitons, we only need to consider up to $n$ virtual deflections of the heavy source at $n$SF. Additional deflections involve extra graviton legs that would either lead to more loops, which would be higher SF, or to contributions outside the potential region. The latter (e.g., at 1SF see bottom right diagram in \figref{fig:TwoLoopFeynRules}) correspond to so-called mushroom diagrams in flat space where a graviton begins and ends on the same matter line.

\subsection{1SF Amplitude at $\mathcal{O}\left(G^2\right)$}
\label{sec:OneSFAmplitude}

We proceed to compute the Compton amplitude, i.e. the amplitude for the emission of two gravitons from the heavy source, and the 1SF amplitude at $\mathcal{O}\left(G^2\right)$ or 2PM. These are the simplest amplitudes where backreaction of the background enters the calculation. 
%

 \begin{figure}
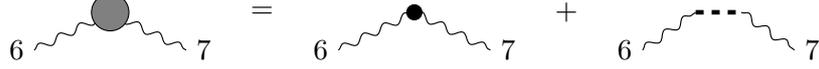

     \centering
     \comptonFeynGraphs
     \caption{The Compton amplitude given as a sum of Feynman diagrams. The gray large blob stands for the on-shell amplitude, while the rest of the symbols are defined in earlier figures. The on-shell gravitons have momenta $q_6$ and $q_7$ and polarization tensors $\varepsilon_6$ and $\varepsilon_7$. We choose this labeling scheme to avoid confusion with subsequent graphs.}
     \label{fig:ComptonFeynGraphs}
 \end{figure}

For the calculation of the Compton amplitude $\mathcal{A}_\text{C}$,  where the subscript `C' stands for `Compton,' there are two diagrams that contribute, which we depict in \figref{fig:ComptonFeynGraphs}.  We refer to the contribution from the first diagram in \figref{fig:ComptonFeynGraphs} as $\mathcal{A}_\text{C}^\text{b}$, where `b' stands for `background insertion,' and the second diagram as $\mathcal{A}_\text{C}^\text{d}$, where `d' stands for `deflection.' We have,
\begin{equation}
\mathcal{A}_\text{C} = \mathcal{A}_\text{C}^\text{b} + \mathcal{A}_\text{C}^\text{d} \,,
\end{equation}
with
\begin{alignat}{2}
\mathcal{A}_\text{C}^\text{b} \, &&=\, &-\frac{8\pi G M}{q_6\cdot q_7} \Bigg(
	\omega^2 (\varepsilon_6 \cdot \varepsilon_7)^2 + \big(
		(v\cdot \varepsilon_7)(\varepsilon_6 \cdot q_7) - (v\cdot \varepsilon_6)(\varepsilon_7 \cdot q_6) \big)^2 \nn \\
	&& &+ 2(\varepsilon_6 \cdot \varepsilon_7) \Big(
	(v\cdot \varepsilon_6) (v\cdot \varepsilon_7)(q_6\cdot q_7) + \omega \big(
		(v\cdot \varepsilon_7)(\varepsilon_6 \cdot q_7) - (v\cdot \varepsilon_6)(\varepsilon_7 \cdot q_6) \big) \Big) \Bigg) \,,\nn \\
\mathcal{A}_\text{C}^\text{d} \, &&= \,  &\frac{8\pi G M}{\omega^2} (v\cdot \varepsilon_6)(v\cdot \varepsilon_7) \Bigg(
		4 \omega^2 (\varepsilon_6 \cdot \varepsilon_7) - (v\cdot \varepsilon_6)(v\cdot \varepsilon_7)(q_6\cdot q_7) \nn\\
		&& &+ 2\omega \big(
		(v\cdot \varepsilon_7)(\varepsilon_6 \cdot q_7) - (v\cdot \varepsilon_6)(\varepsilon_7 \cdot q_6) \big)\Bigg) \,.
  \label{eq:ComptonInParts}
\end{alignat}
In the above we used 
\begin{align}
\omega = v\cdot q_6 =-v\cdot q_7\,, \quad \varepsilon_{6\,\mu\nu} = \varepsilon_{6\,\mu}\, \varepsilon_{6\,\nu} \,, \quad \varepsilon_{7\,\mu\nu} = \varepsilon_{7\,\mu}\,\varepsilon_{7\,\nu}\,,
\end{align}
with $v$ defined below \eqtref{eq:PolyakovActionHeavy}.  Combining the two contributions we arrive at the familiar result,
\begin{align}\label{eq:compton_1}
\mathcal{A}_\text{C} = -\frac{8 \pi G M}{\omega^2 (q_6\cdot q_7)} \Big( &
	\omega^2 (\varepsilon_6 \cdot \varepsilon_7) - (v\cdot \varepsilon_6)(v\cdot \varepsilon_7)(q_6\cdot q_7) \nn \\
	&+ \omega \big(
		(v\cdot \varepsilon_7)(\varepsilon_6 \cdot q_7) - (v\cdot \varepsilon_6)(\varepsilon_7 \cdot q_6) \big)
	 \Big)^2\,,
\end{align}
which manifests the double-copy structure~\cite{Kawai:1985xq,Bern:2008qj,Bern:2010ue} of the Compton amplitude.
All pieces of our Compton amplitude have classical scaling, which would simplify book-keeping of classical contributions especially for higher-order processes.
As mentioned above, to compare the result in Eq.~(\ref{eq:compton_1}) to the corresponding flat-space amplitude, we divide the latter by $2M$ to account for non-relativistic normalization.

In comparing our amplitudes with corresponding flat-space ones, we should be mindful in our mapping of $v$ to the momentum of the heavy particle.
This point, while not relevant for the Compton amplitude itself, may enter in higher-order computations that involve superclassical terms.
Let $p_1$ be the incoming momentum of the heavy particle in flat-space kinematics and $q$ be the total momentum transfer of the process at hand (here $q=q_6+q_7$). Since in curved space we have $v \cdot q = 0$, we may identify
\begin{equation}
\label{eq:pBarDef}
v = \bar{p}_1 / \bar{M} \,, \quad \bar{p}_1 = p_1 + \frac{1}{2} q\,, \quad \bar{M}^2 = \bar{p}_1^2\,.
\end{equation}
This in turn implies that the frame in which we do our curved-space calculations 
%
%
corresponds to the flat-space frame in which $\bar{p}_1^\mu = (\bar{M},0,0,0)$, rather than the one where $p_1^\mu=(M,0,0,0)$. In this frame we have $q^\mu = (0,\bm q)$ and 
\begin{align}
    \gamma = \sigma + \mathcal{O}\big(\bm q^2\big) = y + \mathcal{O}\big(\bm q^2\big) \,,
\end{align}
where $\gamma$ is defined below \eqtref{eq:Leading0SFAmp} and may also be expressed as $\gamma = v\cdot p_2/m$, $\sigma = p_1 \cdot p_2 / (m M)$ and $y = \bar{p}_1 \cdot \bar{p}_2 / (\bar{m} \bar{M})$ with $\bar{p}_2 = p_2 - \frac{1}{2} q$ and $\bar{m}^2 = \bar{p}_2^2$.
Interestingly, we find that the barred variables introduced in Ref.~\cite{Parra-Martinez:2020dzs}, which simplify amplitudes calculations relevant for gravitational-wave physics, naturally appear in our curved-space formalism.

We may make an interesting observation by studying the contribution of the background insertion and that of the deflection separately.  The Compton amplitude is invariant under the transformation
\begin{align}
\varepsilon_6 \rightarrow \varepsilon_6 + \lambda_6 \, q_6 \,, \quad 
\varepsilon_7 \rightarrow \varepsilon_7 + \lambda_7 \, q_7\,,
\end{align}
for any $(\lambda_6,\lambda_7)$. We may think of this as residual gauge freedom at the level of the amplitude and we may use it to simplify its form. In particular, if we choose $(\lambda_6,\lambda_7)$ such that
\begin{align}
\label{eq:polGaugeFix}
v \cdot \varepsilon_6^* = 0 \,, \quad v \cdot \varepsilon_7^* = 0 \,,
\end{align}
where $\varepsilon_{6,7}^*$ signifies the corresponding gauge-fixed polarization, then we have
\begin{align}
\mathcal{A}_\text{C}^\text{d} = 0 \,, \quad
\mathcal{A}_\text{C} = \mathcal{A}_\text{C}^\text{b} = -\frac{8 \pi G M \omega^2}{q_6\cdot q_7} (\varepsilon_6^* \cdot \varepsilon_7^*)^2 \,.
\end{align}
Given this result, one might wonder whether there exists a gauge choice such that the deflection part of the action never contributes, or whether this is an accident due to the simplicity of the particular example.  By inspecting $\actionHFinInt$ in \eqtref{eq:actionHFinInt} we see that, while for all terms linear in $z^\mu$ there is at least one index of $h_{\mu\nu}$ contracted against $v^\mu$, this stops being the case at quadratic order in $z^\mu$. This suggests that the gauge choice in \eqtref{eq:polGaugeFix} would not completely remove the deflection terms when quadratic-in-$z^\mu$ vertices become important. Nevertheless, it would be interesting to understand the extent to which gauge choices of this sort can simplify computations.

 \begin{figure}
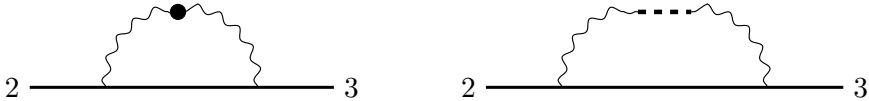

     \centering
     \LeadingGoneSF
     \caption{The Feynman diagrams that contribute to the 1SF amplitude at $\mathcal{O}\left(G^2\right)$. The external probe-particle lines have momenta $p_2$ and $p_3$.}
     \label{fig:LeadingGFeynRules}
 \end{figure}

Finally, we compute the 1SF amplitude at $\mathcal{O}\left(G^2\right)$ via the Feynman diagrams depicted in \figref{fig:LeadingGFeynRules}. We also obtained the result via the generalized unitarity method~\cite{Bern:1994zx,Bern:1994cg}. Our Compton amplitude satisfies the generalized Ward identity, hence we may use the de Donder projector when performing the sewing~\cite{Kosmopoulos:2020pcd}. After some simple integral reduction and integration we find
\begin{align}
\label{eq:oneSFLeadingAmp}
\mathcal{A}_2^\text{1SF, $\mathcal{O}\left(G^2\right)$} = \frac{3 \pi^2 G^2 M m^3 (5\gamma^2-1) }{|q|} \,,
\end{align}
where $\gamma$ and $q$ are defined below \eqsref{eq:Leading0SFAmp}{eq:pBarDef} respectively. This amplitude agrees with the familiar flat-space result~\cite{Bjerrum-Bohr:2002gqz,Cheung:2018wkq} up to non-relativistic normalization. 

The complete $\mathcal{O}\left(G^2\right)$ amplitude is given by the Fourier transform of the second term in \eqtref{eq:PositionSpaceAmpExplicit} together with the one in \eqtref{eq:oneSFLeadingAmp}. While the total amplitude with appropriate normalization is symmetric under $m \leftrightarrow M$ exchange, we see that the contributions at 0SF and 1SF are obtained in quite different ways in our formalism.

\subsection{1SF Amplitude at $\mathcal{O}\left(G^3\right)$}

In this subsection we discuss our computation of the three-point single-graviton-emission amplitude, which corresponds to the flat-space five-point amplitude, and the integrand for the conservative 1SF amplitude at $\mathcal{O}\left(G^3\right)$ or 3PM. 

 \begin{figure}
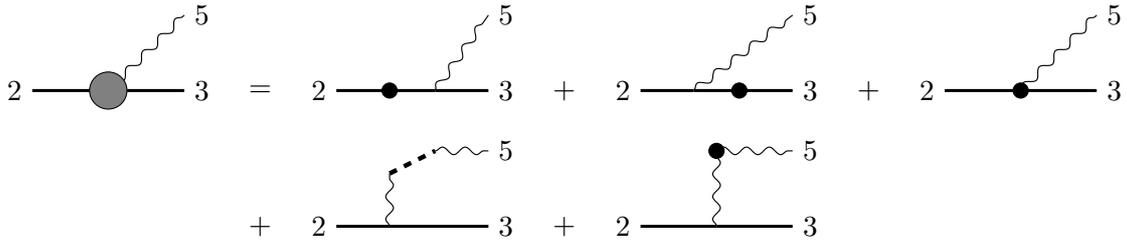

     \centering
     \FivePtFeynGraphs
     \caption{The single-graviton-emission amplitude given as a sum of Feynman diagrams. The gray large blob stands for the on-shell amplitude, while the rest of the symbols are defined in earlier figures. The on-shell graviton has momentum $q_5$ and polarization tensor $\varepsilon_5$. The probe-particle external lines have momenta $p_2$ and $p_3$.}
     \label{fig:FivePtFeynGraphs}
 \end{figure}

The necessary Feynman diagrams for the single-graviton-emission amplitude are shown in \figref{fig:FivePtFeynGraphs}.  While the vertices with background insertions contain all-orders-in-$G$ information, for the case at hand we may truncate them to leading order.
At this order, the vertices arising from $\actionL$ and $\actionLY$ are identical. Similar to the Compton amplitude computed in the previous subsection, we find that for the polarization choice $v \cdot \varepsilon_5^* = 0$ the deflection contributions vanish in the classical limit.

Our amplitude matches the classical limit of the flat-space five-point amplitude obtained in Ref.~\cite{Luna:2017dtq}. Our curved-space amplitude contains superclassical terms associated with the probe particle, which we may keep track of by formally considering complex kinematics. These are contained in the first two diagrams on the right hand side of the equation depicted in \figref{fig:FivePtFeynGraphs}, and match the corresponding terms in the flat-space amplitude. On the other hand, our curved-space amplitudes do not contain superclassical terms associated with the heavy source.  

 \begin{figure}
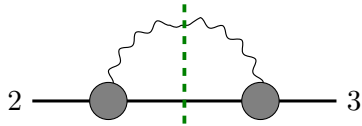

     \centering
     \TwoLoopUnitarity
     \caption{The unitarity cut from which we extract the integrand of the 1SF amplitude at $\mathcal{O}\left(G^3\right)$. The gray large blob stands for the on-shell amplitude we computed above. The vertical green dashed line signifies the unitarity cut, implying that the cut lines are taken on shell. The probe-particle external lines have momenta $p_2$ and $p_3$.}
     \label{fig:TwoLoopUnitarity}
 \end{figure}

We now turn to the 1SF amplitude at $\mathcal{O}\left(G^3\right)$.
We build a generalized-unitarity~\cite{Bern:1994zx,Bern:1994cg} integrand by sewing two single-graviton-emission amplitudes as in \figref{fig:TwoLoopUnitarity}. 
The simplifications of Ref.~\cite{Kosmopoulos:2020pcd} are automatic in this case, given that we are sewing amplitudes with a single graviton.
By taking further cuts we verify that we retrieve the unitarity cuts used to obtain the Hamiltonian at $\mathcal{O}\left(G^3\right)$ in Ref.~\cite{Bern:2019nnu}. In particular, we obtain the so-called N and H cuts, which are relevant for the 1SF part of the amplitude at $\mathcal{O}\left(G^3\right)$. This provides a non-trivial validation of our formalism.

Recently, Ref.~\cite{Adamo:2023cfp} computed the single-graviton-emission amplitude by solving the equations of motion for the probe particle and the graviton in the linearized \sch background. Their analysis employs restricted kinematics where recoil terms do not contribute. 
However, in a generic gauge, the construction of the 1SF amplitude at $\mathcal{O}\left(G^3\right)$ via unitarity would require recoil terms.

 \begin{figure}
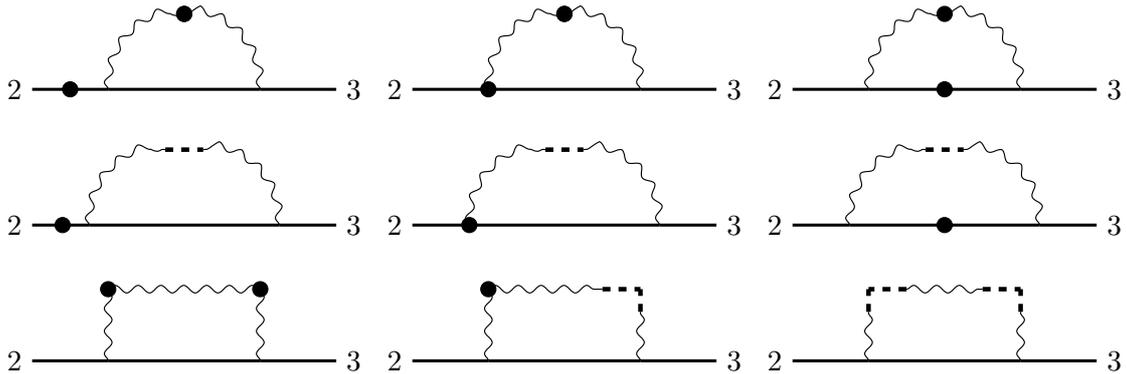

     \centering
     \TwoLoopFeynRules
     \caption{The Feynman diagrams that contribute to the 1SF amplitude at $\mathcal{O}\left(G^3\right)$. The probe-particle external lines have momenta $p_2$ and $p_3$. For the diagrams that are not $2\leftrightarrow 3$ symmetric, one needs to add corresponding ones arising from $2\leftrightarrow 3$ exchange.}
     \label{fig:TwoLoopFeynRules}
 \end{figure}

Alternatively, we also obtain an integrand for the 1SF amplitude at $\mathcal{O}\left(G^3\right)$ using the curved-space Feynman rules. We depict the relevant diagrams in \figref{fig:TwoLoopFeynRules}. As mentioned above, the background vertices may be truncated to linear order in $G$. Also, in accordance with our general discussion in \secref{sec:FeynRulesNSF}, the last diagram in \figref{fig:TwoLoopFeynRules} does not contribute in the potential region. We verify the validity of this integrand in the same way as above.

\section{Conclusions}
\label{sec:Conclusions}
In this paper we used EFT methods and scattering amplitudes in curved space to develop a framework for computing gravitational self-force corrections to binary dynamics, including the effects of the backreaction of the background spacetime. We explicitly verified a number of results, namely the 0SF two-point amplitude to all orders in $G$, the Compton and single-graviton-emission tree-level amplitudes, the 1SF two-point amplitude at $\mathcal{O}\left(G^2\right)$, and the integrand for the 1SF two-point amplitude at $\mathcal{O}\left(G^3\right)$. We find agreement with the literature in all cases.

\begin{figure}
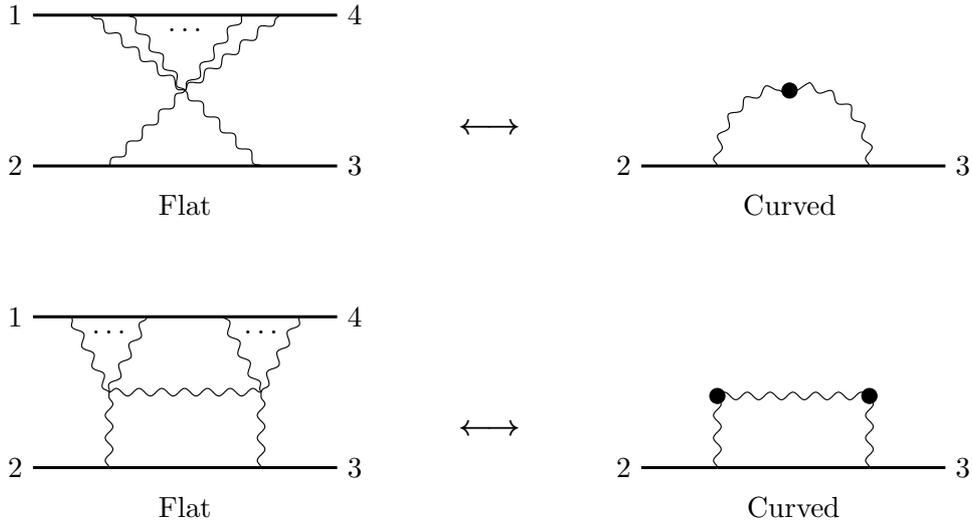

     \centering
     \Hourglass
     \caption{Schematic correspondence between classes of diagrams in a flat-space framework and the curved-space framework of this paper. In the top line, a multi-loop integral on the left is related to a one-loop integral on the right. In the bottom line, a multi-loop integral on the left is related to a two-loop integral on the right. See \secref{sec:FeynRulesNSF} for a discussion of the integral vs diagram loop counting in the curved-space formalism.}
     \label{fig:Hourglass}
 \end{figure}

In our setup, vertices encode information to all orders in $G$ through the background metric, and perturbation theory is organized so that the loop expansion coincides with the SF expansion. Moreover, classes of higher-loop diagrams in flat space are mapped to a few lower-loop diagrams in curved space, which may reduce the number of integrations and automatically incorporate simplifications that would otherwise require non-trivial combinations of diagrams~\cite{Akhoury:2013yua}. We depict an example of this feature in~\figref{fig:Hourglass}, where a family of flat-space diagrams are captured by the background insertion. For instance, a subclass of contributions to the 1SF amplitude and to all orders in $G$ can be obtained as a one-loop computation through the top right of \figref{fig:Hourglass}.

In the future, it would be interesting to consider a Kerr black hole as well as a spinning probe particle. Our formalism can also be applied for computing contributions at higher SF order and at higher order in $G$. For example, one could target the 1SF terms at ${\cal O}(G^5)$ as a first pass at 5PM. 
Radiative effects constitute another natural target; these may be addressed for example by adopting ideas from the amplitudes approach of Ref.~\cite{Kosower:2018adc}, as in, e.g., Refs.~\cite{Herrmann:2021lqe,Herrmann:2021tct}, or from the ``in-in" or ``closed-time-path'' formalism~\cite{Schwinger:1960qe,Keldysh:1964ud,Calzetta:2008iqa}, as in, e.g., Refs.~\cite{Galley:2008ih,Edison:2022cdu,Jakobsen:2022psy,Jakobsen:2023hig,Kalin:2022hph,Dlapa:2022lmu}.
In pursuing such calculations, it would be advantageous to incorporate on-shell methods or to further simplify perturbation theory by choosing special gauges. It would be interesting to see if the advantages of curved-space amplitudes can help with the complexity of calculations at higher PM orders.

Finally, many aspects of our analysis are still anchored to properties of amplitudes in flat space, and it would be interesting to explore alternative methods of calculation that are instead tailored to the inherent properties of amplitudes in curved space. For example, the all-orders-in-$G$ position-space amplitude in \eqtref{eq:Amp0SFPositionSpace} is compact, and suggests that one might calculate in position space where metrics are naturally expressed. Another example is our deference to the flat-space propagator; one might instead consider numerical methods for evaluating Green's functions in curved space.

\section*{Acknowledgments}
We are especially grateful to Carl Beadle, Clifford Cheung, Michael Ruf, Fei Teng and Chia-Hsien Shen. We also wish to thank Zvi Bern, Andrea Cristofoli, Callum Jones, Davide Perrone, Sergio Ricossa, Francesco Riva, Radu Roiban and Francesco Serra. 
The work of D. K. is supported by the Swiss National Science Foundation under grant no. 200021-205016.
The work of M.~S. is supported by the US Department of Energy Early Career program  under award number DE-SC0024224, and the Sloan Foundation.
We are also grateful to the Mani L. Bhaumik Institute for Theoretical Physics for support.
%

\newpage
\bibliography{ref}{}

\providecommand{\href}[2]{#2}\begingroup\raggedright\begin{thebibliography}{100}

\bibitem{LIGOScientific:2016aoc}
{\scshape LIGO Scientific, Virgo} collaboration, \emph{{Observation of
  Gravitational Waves from a Binary Black Hole Merger}},
  \href{https://doi.org/10.1103/PhysRevLett.116.061102}{\emph{Phys. Rev. Lett.}
  {\bfseries 116} (2016) 061102}
  [\href{https://arxiv.org/abs/1602.03837}{{\ttfamily 1602.03837}}].

\bibitem{Buonanno:2022pgc}
A.~Buonanno, M.~Khalil, D.~O'Connell, R.~Roiban, M.P.~Solon and M.~Zeng,
  \emph{{Snowmass White Paper: Gravitational Waves and Scattering Amplitudes}},
   in \emph{{Snowmass 2021}}, 4, 2022
  [\href{https://arxiv.org/abs/2204.05194}{{\ttfamily 2204.05194}}].

\bibitem{Adamo:2022dcm}
T.~Adamo, J.J.M.~Carrasco, M.~Carrillo-Gonz\'alez, M.~Chiodaroli, H.~Elvang,
  H.~Johansson et~al., \emph{{Snowmass White Paper: the Double Copy and its
  Applications}},  in \emph{{Snowmass 2021}}, 4, 2022
  [\href{https://arxiv.org/abs/2204.06547}{{\ttfamily 2204.06547}}].

\bibitem{Bjerrum-Bohr:2022blt}
N.E.J.~Bjerrum-Bohr, P.H.~Damgaard, L.~Plante and P.~Vanhove, \emph{{The SAGEX
  review on scattering amplitudes Chapter 13: Post-Minkowskian expansion from
  scattering amplitudes}},
  \href{https://doi.org/10.1088/1751-8121/ac7a78}{\emph{J. Phys. A} {\bfseries
  55} (2022) 443014} [\href{https://arxiv.org/abs/2203.13024}{{\ttfamily
  2203.13024}}].

\bibitem{Kosower:2022yvp}
D.A.~Kosower, R.~Monteiro and D.~O'Connell, \emph{{The SAGEX review on
  scattering amplitudes Chapter 14: Classical gravity from scattering
  amplitudes}}, \href{https://doi.org/10.1088/1751-8121/ac8846}{\emph{J. Phys.
  A} {\bfseries 55} (2022) 443015}
  [\href{https://arxiv.org/abs/2203.13025}{{\ttfamily 2203.13025}}].

\bibitem{LISA:2017pwj}
{\scshape LISA} collaboration, \emph{{Laser Interferometer Space Antenna}},
  \href{https://arxiv.org/abs/1702.00786}{{\ttfamily 1702.00786}}.

\bibitem{Punturo:2010zz}
M.~Punturo et~al., \emph{{The Einstein Telescope: A third-generation
  gravitational wave observatory}},
  \href{https://doi.org/10.1088/0264-9381/27/19/194002}{\emph{Class. Quant.
  Grav.} {\bfseries 27} (2010) 194002}.

\bibitem{Reitze:2019iox}
D.~Reitze et~al., \emph{{Cosmic Explorer: The U.S. Contribution to
  Gravitational-Wave Astronomy beyond LIGO}}, {\emph{Bull. Am. Astron. Soc.}
  {\bfseries 51} (2019) 035}
  [\href{https://arxiv.org/abs/1907.04833}{{\ttfamily 1907.04833}}].

\bibitem{Kawai:1985xq}
H.~Kawai, D.C.~Lewellen and S.H.H.~Tye, \emph{{A Relation Between Tree
  Amplitudes of Closed and Open Strings}},
  \href{https://doi.org/10.1016/0550-3213(86)90362-7}{\emph{Nucl. Phys. B}
  {\bfseries 269} (1986) 1}.

\bibitem{Bern:2008qj}
Z.~Bern, J.J.M.~Carrasco and H.~Johansson, \emph{{New Relations for
  Gauge-Theory Amplitudes}},
  \href{https://doi.org/10.1103/PhysRevD.78.085011}{\emph{Phys. Rev. D}
  {\bfseries 78} (2008) 085011}
  [\href{https://arxiv.org/abs/0805.3993}{{\ttfamily 0805.3993}}].

\bibitem{Bern:2010ue}
Z.~Bern, J.J.M.~Carrasco and H.~Johansson, \emph{{Perturbative Quantum Gravity
  as a Double Copy of Gauge Theory}},
  \href{https://doi.org/10.1103/PhysRevLett.105.061602}{\emph{Phys. Rev. Lett.}
  {\bfseries 105} (2010) 061602}
  [\href{https://arxiv.org/abs/1004.0476}{{\ttfamily 1004.0476}}].

\bibitem{Bern:2019prr}
Z.~Bern, J.J.~Carrasco, M.~Chiodaroli, H.~Johansson and R.~Roiban, \emph{{The
  Duality Between Color and Kinematics and its Applications}},
  \href{https://arxiv.org/abs/1909.01358}{{\ttfamily 1909.01358}}.

\bibitem{Cheung:2020gbf}
C.~Cheung, N.~Shah and M.P.~Solon, \emph{{Mining the Geodesic Equation for
  Scattering Data}},
  \href{https://doi.org/10.1103/PhysRevD.103.024030}{\emph{Phys. Rev. D}
  {\bfseries 103} (2021) 024030}
  [\href{https://arxiv.org/abs/2010.08568}{{\ttfamily 2010.08568}}].

\bibitem{Cristofoli:2021jas}
A.~Cristofoli, R.~Gonzo, N.~Moynihan, D.~O'Connell, A.~Ross, M.~Sergola et~al.,
  \emph{{The Uncertainty Principle and Classical Amplitudes}},
  \href{https://arxiv.org/abs/2112.07556}{{\ttfamily 2112.07556}}.

\bibitem{Bern:2021dqo}
Z.~Bern, J.~Parra-Martinez, R.~Roiban, M.S.~Ruf, C.-H.~Shen, M.P.~Solon et~al.,
  \emph{{Scattering Amplitudes and Conservative Binary Dynamics at ${\cal
  O}(G^4)$}}, \href{https://doi.org/10.1103/PhysRevLett.126.171601}{\emph{Phys.
  Rev. Lett.} {\bfseries 126} (2021) 171601}
  [\href{https://arxiv.org/abs/2101.07254}{{\ttfamily 2101.07254}}].

\bibitem{Blanchet:2013haa}
L.~Blanchet, \emph{{Gravitational Radiation from Post-Newtonian Sources and
  Inspiralling Compact Binaries}},
  \href{https://doi.org/10.12942/lrr-2014-2}{\emph{Living Rev. Rel.} {\bfseries
  17} (2014) 2} [\href{https://arxiv.org/abs/1310.1528}{{\ttfamily
  1310.1528}}].

\bibitem{Goldberger:2004jt}
W.D.~Goldberger and I.Z.~Rothstein, \emph{{An Effective field theory of gravity
  for extended objects}},
  \href{https://doi.org/10.1103/PhysRevD.73.104029}{\emph{Phys. Rev. D}
  {\bfseries 73} (2006) 104029}
  [\href{https://arxiv.org/abs/hep-th/0409156}{{\ttfamily hep-th/0409156}}].

\bibitem{Porto:2016pyg}
R.A.~Porto, \emph{{The effective field theorist\textquoteright{}s approach to
  gravitational dynamics}},
  \href{https://doi.org/10.1016/j.physrep.2016.04.003}{\emph{Phys. Rept.}
  {\bfseries 633} (2016) 1} [\href{https://arxiv.org/abs/1601.04914}{{\ttfamily
  1601.04914}}].

\bibitem{Levi:2018nxp}
M.~Levi, \emph{{Effective Field Theories of Post-Newtonian Gravity: A
  comprehensive review}},
  \href{https://doi.org/10.1088/1361-6633/ab12bc}{\emph{Rept. Prog. Phys.}
  {\bfseries 83} (2020) 075901}
  [\href{https://arxiv.org/abs/1807.01699}{{\ttfamily 1807.01699}}].

\bibitem{Bern:2021yeh}
Z.~Bern, J.~Parra-Martinez, R.~Roiban, M.S.~Ruf, C.-H.~Shen, M.P.~Solon et~al.,
  \emph{{Scattering Amplitudes, the Tail Effect, and Conservative Binary
  Dynamics at O(G4)}},
  \href{https://doi.org/10.1103/PhysRevLett.128.161103}{\emph{Phys. Rev. Lett.}
  {\bfseries 128} (2022) 161103}
  [\href{https://arxiv.org/abs/2112.10750}{{\ttfamily 2112.10750}}].

\bibitem{Dlapa:2021npj}
C.~Dlapa, G.~K\"alin, Z.~Liu and R.A.~Porto, \emph{{Dynamics of Binary Systems
  to Fourth Post-Minkowskian Order from the Effective Field Theory Approach}},
  \href{https://arxiv.org/abs/2106.08276}{{\ttfamily 2106.08276}}.

\bibitem{Dlapa:2021vgp}
C.~Dlapa, G.~K\"alin, Z.~Liu and R.A.~Porto, \emph{{Conservative Dynamics of
  Binary Systems at Fourth Post-Minkowskian Order in the Large-Eccentricity
  Expansion}},
  \href{https://doi.org/10.1103/PhysRevLett.128.161104}{\emph{Phys. Rev. Lett.}
  {\bfseries 128} (2022) 161104}
  [\href{https://arxiv.org/abs/2112.11296}{{\ttfamily 2112.11296}}].

\bibitem{Dlapa:2022lmu}
C.~Dlapa, G.~K\"alin, Z.~Liu, J.~Neef and R.A.~Porto, \emph{{Radiation Reaction
  and Gravitational Waves at Fourth Post-Minkowskian Order}},
  \href{https://doi.org/10.1103/PhysRevLett.130.101401}{\emph{Phys. Rev. Lett.}
  {\bfseries 130} (2023) 101401}
  [\href{https://arxiv.org/abs/2210.05541}{{\ttfamily 2210.05541}}].

\bibitem{Damgaard:2023ttc}
P.H.~Damgaard, E.R.~Hansen, L.~Plant\'e and P.~Vanhove, \emph{{Classical
  Observables from the Exponential Representation of the Gravitational
  S-Matrix}},  \href{https://arxiv.org/abs/2307.04746}{{\ttfamily 2307.04746}}.

\bibitem{Jakobsen:2023ndj}
G.U.~Jakobsen, G.~Mogull, J.~Plefka, B.~Sauer and Y.~Xu, \emph{{Conservative
  scattering of spinning black holes at fourth post-Minkowskian order}},
  \href{https://arxiv.org/abs/2306.01714}{{\ttfamily 2306.01714}}.

\bibitem{Mino:1996nk}
Y.~Mino, M.~Sasaki and T.~Tanaka, \emph{{Gravitational radiation reaction to a
  particle motion}},
  \href{https://doi.org/10.1103/PhysRevD.55.3457}{\emph{Phys. Rev. D}
  {\bfseries 55} (1997) 3457}
  [\href{https://arxiv.org/abs/gr-qc/9606018}{{\ttfamily gr-qc/9606018}}].

\bibitem{Quinn:1996am}
T.C.~Quinn and R.M.~Wald, \emph{{An Axiomatic approach to electromagnetic and
  gravitational radiation reaction of particles in curved space-time}},
  \href{https://doi.org/10.1103/PhysRevD.56.3381}{\emph{Phys. Rev. D}
  {\bfseries 56} (1997) 3381}
  [\href{https://arxiv.org/abs/gr-qc/9610053}{{\ttfamily gr-qc/9610053}}].

\bibitem{Detweiler:2000gt}
S.L.~Detweiler, \emph{{Radiation reaction and the selfforce for a point mass in
  general relativity}},
  \href{https://doi.org/10.1103/PhysRevLett.86.1931}{\emph{Phys. Rev. Lett.}
  {\bfseries 86} (2001) 1931}
  [\href{https://arxiv.org/abs/gr-qc/0011039}{{\ttfamily gr-qc/0011039}}].

\bibitem{Detweiler:2002mi}
S.L.~Detweiler and B.F.~Whiting, \emph{{Selfforce via a Green's function
  decomposition}},
  \href{https://doi.org/10.1103/PhysRevD.67.024025}{\emph{Phys. Rev. D}
  {\bfseries 67} (2003) 024025}
  [\href{https://arxiv.org/abs/gr-qc/0202086}{{\ttfamily gr-qc/0202086}}].

\bibitem{Gralla:2008fg}
S.E.~Gralla and R.M.~Wald, \emph{{A Rigorous Derivation of Gravitational
  Self-force}},
  \href{https://doi.org/10.1088/0264-9381/25/20/205009}{\emph{Class. Quant.
  Grav.} {\bfseries 25} (2008) 205009}
  [\href{https://arxiv.org/abs/0806.3293}{{\ttfamily 0806.3293}}].

\bibitem{Pound:2009sm}
A.~Pound, \emph{{Self-consistent gravitational self-force}},
  \href{https://doi.org/10.1103/PhysRevD.81.024023}{\emph{Phys. Rev. D}
  {\bfseries 81} (2010) 024023}
  [\href{https://arxiv.org/abs/0907.5197}{{\ttfamily 0907.5197}}].

\bibitem{Rosenthal:2006iy}
E.~Rosenthal, \emph{{Second-order gravitational self-force}},
  \href{https://doi.org/10.1103/PhysRevD.74.084018}{\emph{Phys. Rev. D}
  {\bfseries 74} (2006) 084018}
  [\href{https://arxiv.org/abs/gr-qc/0609069}{{\ttfamily gr-qc/0609069}}].

\bibitem{Detweiler:2011tt}
S.~Detweiler, \emph{{Gravitational radiation reaction and second order
  perturbation theory}},
  \href{https://doi.org/10.1103/PhysRevD.85.044048}{\emph{Phys. Rev. D}
  {\bfseries 85} (2012) 044048}
  [\href{https://arxiv.org/abs/1107.2098}{{\ttfamily 1107.2098}}].

\bibitem{Pound:2012nt}
A.~Pound, \emph{{Second-order gravitational self-force}},
  \href{https://doi.org/10.1103/PhysRevLett.109.051101}{\emph{Phys. Rev. Lett.}
  {\bfseries 109} (2012) 051101}
  [\href{https://arxiv.org/abs/1201.5089}{{\ttfamily 1201.5089}}].

\bibitem{Gralla:2012db}
S.E.~Gralla, \emph{{Second Order Gravitational Self Force}},
  \href{https://doi.org/10.1103/PhysRevD.85.124011}{\emph{Phys. Rev. D}
  {\bfseries 85} (2012) 124011}
  [\href{https://arxiv.org/abs/1203.3189}{{\ttfamily 1203.3189}}].

\bibitem{Barack:2018yvs}
L.~Barack and A.~Pound, \emph{{Self-force and radiation reaction in general
  relativity}}, \href{https://doi.org/10.1088/1361-6633/aae552}{\emph{Rept.
  Prog. Phys.} {\bfseries 82} (2019) 016904}
  [\href{https://arxiv.org/abs/1805.10385}{{\ttfamily 1805.10385}}].

\bibitem{Pound:2021qin}
A.~Pound and B.~Wardell, \emph{{Black hole perturbation theory and
  gravitational self-force}},
  \href{https://arxiv.org/abs/2101.04592}{{\ttfamily 2101.04592}}.

\bibitem{Galley:2006gs}
C.R.~Galley, B.L.~Hu and S.-Y.~Lin, \emph{{Electromagnetic and gravitational
  self-force on a relativistic particle from quantum fields in curved space}},
  \href{https://doi.org/10.1103/PhysRevD.74.024017}{\emph{Phys. Rev. D}
  {\bfseries 74} (2006) 024017}
  [\href{https://arxiv.org/abs/gr-qc/0603099}{{\ttfamily gr-qc/0603099}}].

\bibitem{Galley:2008ih}
C.R.~Galley and B.L.~Hu, \emph{{Self-force on extreme mass ratio inspirals via
  curved spacetime effective field theory}},
  \href{https://doi.org/10.1103/PhysRevD.79.064002}{\emph{Phys. Rev. D}
  {\bfseries 79} (2009) 064002}
  [\href{https://arxiv.org/abs/0801.0900}{{\ttfamily 0801.0900}}].

\bibitem{Hinderer:2008dm}
T.~Hinderer and E.E.~Flanagan, \emph{{Two timescale analysis of extreme mass
  ratio inspirals in Kerr. I. Orbital Motion}},
  \href{https://doi.org/10.1103/PhysRevD.78.064028}{\emph{Phys. Rev. D}
  {\bfseries 78} (2008) 064028}
  [\href{https://arxiv.org/abs/0805.3337}{{\ttfamily 0805.3337}}].

\bibitem{Isoyama:2012bx}
S.~Isoyama, R.~Fujita, N.~Sago, H.~Tagoshi and T.~Tanaka, \emph{{Impact of the
  second-order self-forces on the dephasing of the gravitational waves from
  quasicircular extreme mass-ratio inspirals}},
  \href{https://doi.org/10.1103/PhysRevD.87.024010}{\emph{Phys. Rev. D}
  {\bfseries 87} (2013) 024010}
  [\href{https://arxiv.org/abs/1210.2569}{{\ttfamily 1210.2569}}].

\bibitem{Burko:2013cca}
L.M.~Burko and G.~Khanna, \emph{{Self-force gravitational waveforms for extreme
  and intermediate mass ratio inspirals. II: Importance of the second-order
  dissipative effect}},
  \href{https://doi.org/10.1103/PhysRevD.88.024002}{\emph{Phys. Rev. D}
  {\bfseries 88} (2013) 024002}
  [\href{https://arxiv.org/abs/1304.5296}{{\ttfamily 1304.5296}}].

\bibitem{Barack:2010tm}
L.~Barack and N.~Sago, \emph{{Gravitational self-force on a particle in
  eccentric orbit around a Schwarzschild black hole}},
  \href{https://doi.org/10.1103/PhysRevD.81.084021}{\emph{Phys. Rev. D}
  {\bfseries 81} (2010) 084021}
  [\href{https://arxiv.org/abs/1002.2386}{{\ttfamily 1002.2386}}].

\bibitem{Barack:2011ed}
L.~Barack and N.~Sago, \emph{{Beyond the geodesic approximation: conservative
  effects of the gravitational self-force in eccentric orbits around a
  Schwarzschild black hole}},
  \href{https://doi.org/10.1103/PhysRevD.83.084023}{\emph{Phys. Rev. D}
  {\bfseries 83} (2011) 084023}
  [\href{https://arxiv.org/abs/1101.3331}{{\ttfamily 1101.3331}}].

\bibitem{Wardell:2021fyy}
B.~Wardell, A.~Pound, N.~Warburton, J.~Miller, L.~Durkan and A.~Le~Tiec,
  \emph{{Gravitational Waveforms for Compact Binaries from Second-Order
  Self-Force Theory}},
  \href{https://doi.org/10.1103/PhysRevLett.130.241402}{\emph{Phys. Rev. Lett.}
  {\bfseries 130} (2023) 241402}
  [\href{https://arxiv.org/abs/2112.12265}{{\ttfamily 2112.12265}}].

\bibitem{Gralla:2021qaf}
S.E.~Gralla and K.~Lobo, \emph{{Self-force effects in post-Minkowskian
  scattering}}, \href{https://doi.org/10.1088/1361-6382/ac5d88}{\emph{Class.
  Quant. Grav.} {\bfseries 39} (2022) 095001}
  [\href{https://arxiv.org/abs/2110.08681}{{\ttfamily 2110.08681}}].

\bibitem{Barack:2022pde}
L.~Barack and O.~Long, \emph{{Self-force correction to the deflection angle in
  black-hole scattering: A scalar charge toy model}},
  \href{https://doi.org/10.1103/PhysRevD.106.104031}{\emph{Phys. Rev. D}
  {\bfseries 106} (2022) 104031}
  [\href{https://arxiv.org/abs/2209.03740}{{\ttfamily 2209.03740}}].

\bibitem{Barack:2023oqp}
L.~Barack et~al., \emph{{Comparison of post-Minkowskian and self-force
  expansions: Scattering in a scalar charge toy model}},
  \href{https://doi.org/10.1103/PhysRevD.108.024025}{\emph{Phys. Rev. D}
  {\bfseries 108} (2023) 024025}
  [\href{https://arxiv.org/abs/2304.09200}{{\ttfamily 2304.09200}}].

\bibitem{Duff:1973zz}
M.J.~Duff, \emph{{Quantum Tree Graphs and the Schwarzschild Solution}},
  \href{https://doi.org/10.1103/PhysRevD.7.2317}{\emph{Phys. Rev. D} {\bfseries
  7} (1973) 2317}.

\bibitem{Neill:2013wsa}
D.~Neill and I.Z.~Rothstein, \emph{{Classical Space-Times from the S Matrix}},
  \href{https://doi.org/10.1016/j.nuclphysb.2013.09.007}{\emph{Nucl. Phys. B}
  {\bfseries 877} (2013) 177}
  [\href{https://arxiv.org/abs/1304.7263}{{\ttfamily 1304.7263}}].

\bibitem{Mougiakakos:2020laz}
S.~Mougiakakos and P.~Vanhove, \emph{{Schwarzschild-Tangherlini metric from
  scattering amplitudes in various dimensions}},
  \href{https://doi.org/10.1103/PhysRevD.103.026001}{\emph{Phys. Rev. D}
  {\bfseries 103} (2021) 026001}
  [\href{https://arxiv.org/abs/2010.08882}{{\ttfamily 2010.08882}}].

\bibitem{Jakobsen:2020ksu}
G.U.~Jakobsen, \emph{{Schwarzschild-Tangherlini Metric from Scattering
  Amplitudes}}, \href{https://doi.org/10.1103/PhysRevD.102.104065}{\emph{Phys.
  Rev. D} {\bfseries 102} (2020) 104065}
  [\href{https://arxiv.org/abs/2006.01734}{{\ttfamily 2006.01734}}].

\bibitem{DOnofrio:2022cvn}
S.~D'Onofrio, F.~Fragomeno, C.~Gambino and F.~Riccioni, \emph{{The
  Reissner-Nordstr\"om-Tangherlini solution from scattering amplitudes of
  charged scalars}}, \href{https://doi.org/10.1007/JHEP09(2022)013}{\emph{JHEP}
  {\bfseries 09} (2022) 013}
  [\href{https://arxiv.org/abs/2207.05841}{{\ttfamily 2207.05841}}].

\bibitem{Siemonsen:2019dsu}
N.~Siemonsen and J.~Vines, \emph{{Test black holes, scattering amplitudes and
  perturbations of Kerr spacetime}},
  \href{https://doi.org/10.1103/PhysRevD.101.064066}{\emph{Phys. Rev. D}
  {\bfseries 101} (2020) 064066}
  [\href{https://arxiv.org/abs/1909.07361}{{\ttfamily 1909.07361}}].

\bibitem{Guevara:2019fsj}
A.~Guevara, A.~Ochirov and J.~Vines, \emph{{Black-hole scattering with general
  spin directions from minimal-coupling amplitudes}},
  \href{https://doi.org/10.1103/PhysRevD.100.104024}{\emph{Phys. Rev. D}
  {\bfseries 100} (2019) 104024}
  [\href{https://arxiv.org/abs/1906.10071}{{\ttfamily 1906.10071}}].

\bibitem{Guevara:2018wpp}
A.~Guevara, A.~Ochirov and J.~Vines, \emph{{Scattering of Spinning Black Holes
  from Exponentiated Soft Factors}},
  \href{https://doi.org/10.1007/JHEP09(2019)056}{\emph{JHEP} {\bfseries 09}
  (2019) 056} [\href{https://arxiv.org/abs/1812.06895}{{\ttfamily
  1812.06895}}].

\bibitem{Chung:2018kqs}
M.-Z.~Chung, Y.-T.~Huang, J.-W.~Kim and S.~Lee, \emph{{The simplest massive
  S-matrix: from minimal coupling to Black Holes}},
  \href{https://doi.org/10.1007/JHEP04(2019)156}{\emph{JHEP} {\bfseries 04}
  (2019) 156} [\href{https://arxiv.org/abs/1812.08752}{{\ttfamily
  1812.08752}}].

\bibitem{Chung:2019duq}
M.-Z.~Chung, Y.-T.~Huang and J.-W.~Kim, \emph{{Classical potential for general
  spinning bodies}}, \href{https://doi.org/10.1007/JHEP09(2020)074}{\emph{JHEP}
  {\bfseries 09} (2020) 074}
  [\href{https://arxiv.org/abs/1908.08463}{{\ttfamily 1908.08463}}].

\bibitem{Chung:2020rrz}
M.-Z.~Chung, Y.-t.~Huang, J.-W.~Kim and S.~Lee, \emph{{Complete Hamiltonian for
  spinning binary systems at first post-Minkowskian order}},
  \href{https://doi.org/10.1007/JHEP05(2020)105}{\emph{JHEP} {\bfseries 05}
  (2020) 105} [\href{https://arxiv.org/abs/2003.06600}{{\ttfamily
  2003.06600}}].

\bibitem{Chen:2021kxt}
W.-M.~Chen, M.-Z.~Chung, Y.-t.~Huang and J.-W.~Kim, \emph{{The 2PM Hamiltonian
  for binary Kerr to quartic in spin}},
  \href{https://doi.org/10.1007/JHEP08(2022)148}{\emph{JHEP} {\bfseries 08}
  (2022) 148} [\href{https://arxiv.org/abs/2111.13639}{{\ttfamily
  2111.13639}}].

\bibitem{Arkani-Hamed:2019ymq}
N.~Arkani-Hamed, Y.-t.~Huang and D.~O'Connell, \emph{{Kerr black holes as
  elementary particles}},
  \href{https://doi.org/10.1007/JHEP01(2020)046}{\emph{JHEP} {\bfseries 01}
  (2020) 046} [\href{https://arxiv.org/abs/1906.10100}{{\ttfamily
  1906.10100}}].

\bibitem{Bern:2020buy}
Z.~Bern, A.~Luna, R.~Roiban, C.-H.~Shen and M.~Zeng, \emph{{Spinning black hole
  binary dynamics, scattering amplitudes, and effective field theory}},
  \href{https://doi.org/10.1103/PhysRevD.104.065014}{\emph{Phys. Rev. D}
  {\bfseries 104} (2021) 065014}
  [\href{https://arxiv.org/abs/2005.03071}{{\ttfamily 2005.03071}}].

\bibitem{Kosmopoulos:2021zoq}
D.~Kosmopoulos and A.~Luna, \emph{{Quadratic-in-spin Hamiltonian at $
  \mathcal{O} $(G$^{2}$) from scattering amplitudes}},
  \href{https://doi.org/10.1007/JHEP07(2021)037}{\emph{JHEP} {\bfseries 07}
  (2021) 037} [\href{https://arxiv.org/abs/2102.10137}{{\ttfamily
  2102.10137}}].

\bibitem{Bern:2022kto}
Z.~Bern, D.~Kosmopoulos, A.~Luna, R.~Roiban and F.~Teng, \emph{{Binary Dynamics
  through the Fifth Power of Spin at O(G2)}},
  \href{https://doi.org/10.1103/PhysRevLett.130.201402}{\emph{Phys. Rev. Lett.}
  {\bfseries 130} (2023) 201402}
  [\href{https://arxiv.org/abs/2203.06202}{{\ttfamily 2203.06202}}].

\bibitem{Aoude:2020onz}
R.~Aoude, K.~Haddad and A.~Helset, \emph{{On-shell heavy particle effective
  theories}}, \href{https://doi.org/10.1007/JHEP05(2020)051}{\emph{JHEP}
  {\bfseries 05} (2020) 051}
  [\href{https://arxiv.org/abs/2001.09164}{{\ttfamily 2001.09164}}].

\bibitem{Aoude:2022trd}
R.~Aoude, K.~Haddad and A.~Helset, \emph{{Searching for Kerr in the 2PM
  amplitude}},  \href{https://arxiv.org/abs/2203.06197}{{\ttfamily
  2203.06197}}.

\bibitem{Aoude:2023vdk}
R.~Aoude, K.~Haddad and A.~Helset, \emph{{Classical gravitational scattering
  amplitude at O(G2S1\ensuremath{\infty}S2\ensuremath{\infty})}},
  \href{https://doi.org/10.1103/PhysRevD.108.024050}{\emph{Phys. Rev. D}
  {\bfseries 108} (2023) 024050}
  [\href{https://arxiv.org/abs/2304.13740}{{\ttfamily 2304.13740}}].

\bibitem{Aoude:2022thd}
R.~Aoude, K.~Haddad and A.~Helset, \emph{{Classical Gravitational
  Spinning-Spinless Scattering at O(G2S\ensuremath{\infty})}},
  \href{https://doi.org/10.1103/PhysRevLett.129.141102}{\emph{Phys. Rev. Lett.}
  {\bfseries 129} (2022) 141102}
  [\href{https://arxiv.org/abs/2205.02809}{{\ttfamily 2205.02809}}].

\bibitem{Maybee:2019jus}
B.~Maybee, D.~O'Connell and J.~Vines, \emph{{Observables and amplitudes for
  spinning particles and black holes}},
  \href{https://doi.org/10.1007/JHEP12(2019)156}{\emph{JHEP} {\bfseries 12}
  (2019) 156} [\href{https://arxiv.org/abs/1906.09260}{{\ttfamily
  1906.09260}}].

\bibitem{Bjerrum-Bohr:2023jau}
N.E.J.~Bjerrum-Bohr, G.~Chen and M.~Skowronek, \emph{{Classical spin
  gravitational Compton scattering}},
  \href{https://doi.org/10.1007/JHEP06(2023)170}{\emph{JHEP} {\bfseries 06}
  (2023) 170} [\href{https://arxiv.org/abs/2302.00498}{{\ttfamily
  2302.00498}}].

\bibitem{Adamo:2020qru}
T.~Adamo and A.~Ilderton, \emph{{Classical and quantum double copy of
  back-reaction}}, \href{https://doi.org/10.1007/JHEP09(2020)200}{\emph{JHEP}
  {\bfseries 09} (2020) 200}
  [\href{https://arxiv.org/abs/2005.05807}{{\ttfamily 2005.05807}}].

\bibitem{Cristofoli:2020hnk}
A.~Cristofoli, \emph{{Gravitational shock waves and scattering amplitudes}},
  \href{https://doi.org/10.1007/JHEP11(2020)160}{\emph{JHEP} {\bfseries 11}
  (2020) 160} [\href{https://arxiv.org/abs/2006.08283}{{\ttfamily
  2006.08283}}].

\bibitem{Adamo:2021rfq}
T.~Adamo, A.~Cristofoli and P.~Tourkine, \emph{{Eikonal amplitudes from curved
  backgrounds}},
  \href{https://doi.org/10.21468/SciPostPhys.13.2.032}{\emph{SciPost Phys.}
  {\bfseries 13} (2022) 032}
  [\href{https://arxiv.org/abs/2112.09113}{{\ttfamily 2112.09113}}].

\bibitem{Adamo:2022mev}
T.~Adamo, L.~Mason and A.~Sharma, \emph{{Graviton scattering in self-dual
  radiative space-times}},
  \href{https://doi.org/10.1088/1361-6382/acc233}{\emph{Class. Quant. Grav.}
  {\bfseries 40} (2023) 095002}
  [\href{https://arxiv.org/abs/2203.02238}{{\ttfamily 2203.02238}}].

\bibitem{Adamo:2022rmp}
T.~Adamo, A.~Cristofoli and A.~Ilderton, \emph{{Classical physics from
  amplitudes on curved backgrounds}},
  \href{https://doi.org/10.1007/JHEP08(2022)281}{\emph{JHEP} {\bfseries 08}
  (2022) 281} [\href{https://arxiv.org/abs/2203.13785}{{\ttfamily
  2203.13785}}].

\bibitem{Adamo:2022qci}
T.~Adamo, A.~Cristofoli, A.~Ilderton and S.~Klisch, \emph{{All Order
  Gravitational Waveforms from Scattering Amplitudes}},
  \href{https://doi.org/10.1103/PhysRevLett.131.011601}{\emph{Phys. Rev. Lett.}
  {\bfseries 131} (2023) 011601}
  [\href{https://arxiv.org/abs/2210.04696}{{\ttfamily 2210.04696}}].

\bibitem{Adamo:2023cfp}
T.~Adamo, A.~Cristofoli, A.~Ilderton and S.~Klisch, \emph{{Scattering
  amplitudes for self-force}},
  \href{https://arxiv.org/abs/2307.00431}{{\ttfamily 2307.00431}}.

\bibitem{Low:2001bw}
I.~Low and A.V.~Manohar, \emph{{Spontaneously broken space-time symmetries and
  Goldstone's theorem}},
  \href{https://doi.org/10.1103/PhysRevLett.88.101602}{\emph{Phys. Rev. Lett.}
  {\bfseries 88} (2002) 101602}
  [\href{https://arxiv.org/abs/hep-th/0110285}{{\ttfamily hep-th/0110285}}].

\bibitem{Watanabe:2013iia}
H.~Watanabe and H.~Murayama, \emph{{Redundancies in Nambu-Goldstone Bosons}},
  \href{https://doi.org/10.1103/PhysRevLett.110.181601}{\emph{Phys. Rev. Lett.}
  {\bfseries 110} (2013) 181601}
  [\href{https://arxiv.org/abs/1302.4800}{{\ttfamily 1302.4800}}].

\bibitem{Delacretaz:2014oxa}
L.V.~Delacr\'etaz, S.~Endlich, A.~Monin, R.~Penco and F.~Riva,
  \emph{{(Re-)Inventing the Relativistic Wheel: Gravity, Cosets, and Spinning
  Objects}}, \href{https://doi.org/10.1007/JHEP11(2014)008}{\emph{JHEP}
  {\bfseries 11} (2014) 008} [\href{https://arxiv.org/abs/1405.7384}{{\ttfamily
  1405.7384}}].

\bibitem{Bjerrum-Bohr:2002gqz}
N.E.J.~Bjerrum-Bohr, J.F.~Donoghue and B.R.~Holstein, \emph{{Quantum
  gravitational corrections to the nonrelativistic scattering potential of two
  masses}}, \href{https://doi.org/10.1103/PhysRevD.71.069903}{\emph{Phys. Rev.
  D} {\bfseries 67} (2003) 084033}
  [\href{https://arxiv.org/abs/hep-th/0211072}{{\ttfamily hep-th/0211072}}].

\bibitem{Bern:2019nnu}
Z.~Bern, C.~Cheung, R.~Roiban, C.-H.~Shen, M.P.~Solon and M.~Zeng,
  \emph{{Scattering Amplitudes and the Conservative Hamiltonian for Binary
  Systems at Third Post-Minkowskian Order}},
  \href{https://doi.org/10.1103/PhysRevLett.122.201603}{\emph{Phys. Rev. Lett.}
  {\bfseries 122} (2019) 201603}
  [\href{https://arxiv.org/abs/1901.04424}{{\ttfamily 1901.04424}}].

\bibitem{Luna:2017dtq}
A.~Luna, I.~Nicholson, D.~O'Connell and C.D.~White, \emph{{Inelastic Black Hole
  Scattering from Charged Scalar Amplitudes}},
  \href{https://doi.org/10.1007/JHEP03(2018)044}{\emph{JHEP} {\bfseries 03}
  (2018) 044} [\href{https://arxiv.org/abs/1711.03901}{{\ttfamily
  1711.03901}}].

\bibitem{Cheung:2023lnj}
C.~Cheung, J.~Parra-Martinez, I.Z.~Rothstein, N.~Shah and J.~Wilson-Gerow,
  \emph{{Effective Field Theory for Extreme Mass Ratios}},
  \href{https://arxiv.org/abs/2308.14832}{{\ttfamily 2308.14832}}.

\bibitem{Birrell:1982ix}
N.D.~Birrell and P.C.W.~Davies, \emph{{Quantum Fields in Curved Space}},
  Cambridge Monographs on Mathematical Physics, Cambridge Univ. Press,
  Cambridge, UK (2, 1984),
  \href{https://doi.org/10.1017/CBO9780511622632}{10.1017/CBO9780511622632}.

\bibitem{Parker:2009uva}
L.E.~Parker and D.~Toms, \emph{{Quantum Field Theory in Curved Spacetime}:
  {Quantized Field and Gravity}}, Cambridge Monographs on Mathematical Physics,
  Cambridge University Press (8, 2009),
  \href{https://doi.org/10.1017/CBO9780511813924}{10.1017/CBO9780511813924}.

\bibitem{Wald:1975kc}
R.M.~Wald, \emph{{On Particle Creation by Black Holes}},
  \href{https://doi.org/10.1007/BF01609863}{\emph{Commun. Math. Phys.}
  {\bfseries 45} (1975) 9}.

\bibitem{Parker:1975jm}
L.~Parker, \emph{{Probability Distribution of Particles Created by a Black
  Hole}}, \href{https://doi.org/10.1103/PhysRevD.12.1519}{\emph{Phys. Rev. D}
  {\bfseries 12} (1975) 1519}.

\bibitem{Hawking:1975vcx}
S.W.~Hawking, \emph{{Particle Creation by Black Holes}},
  \href{https://doi.org/10.1007/BF02345020}{\emph{Commun. Math. Phys.}
  {\bfseries 43} (1975) 199}.

\bibitem{Tangherlini:1963bw}
F.R.~Tangherlini, \emph{{Schwarzschild field in n dimensions and the
  dimensionality of space problem}},
  \href{https://doi.org/10.1007/BF02784569}{\emph{Nuovo Cim.} {\bfseries 27}
  (1963) 636}.

\bibitem{Komissarov:2022gax}
I.~Komissarov, A.~Nicolis and J.~Staunton, \emph{{Cosmology as a weak
  gravitational field and the trans-Planckian problem}},
  \href{https://doi.org/10.1007/JHEP05(2023)216}{\emph{JHEP} {\bfseries 05}
  (2023) 216} [\href{https://arxiv.org/abs/2210.11508}{{\ttfamily
  2210.11508}}].

\bibitem{Cheung:2018wkq}
C.~Cheung, I.Z.~Rothstein and M.P.~Solon, \emph{{From Scattering Amplitudes to
  Classical Potentials in the Post-Minkowskian Expansion}},
  \href{https://doi.org/10.1103/PhysRevLett.121.251101}{\emph{Phys. Rev. Lett.}
  {\bfseries 121} (2018) 251101}
  [\href{https://arxiv.org/abs/1808.02489}{{\ttfamily 1808.02489}}].

\bibitem{Kalin:2019rwq}
G.~K\"alin and R.A.~Porto, \emph{{From Boundary Data to Bound States}},
  \href{https://doi.org/10.1007/JHEP01(2020)072}{\emph{JHEP} {\bfseries 01}
  (2020) 072} [\href{https://arxiv.org/abs/1910.03008}{{\ttfamily
  1910.03008}}].

\bibitem{Bern:2019crd}
Z.~Bern, C.~Cheung, R.~Roiban, C.-H.~Shen, M.P.~Solon and M.~Zeng, \emph{{Black
  Hole Binary Dynamics from the Double Copy and Effective Theory}},
  \href{https://doi.org/10.1007/JHEP10(2019)206}{\emph{JHEP} {\bfseries 10}
  (2019) 206} [\href{https://arxiv.org/abs/1908.01493}{{\ttfamily
  1908.01493}}].

\bibitem{Wise:1992hn}
M.B.~Wise, \emph{{Chiral perturbation theory for hadrons containing a heavy
  quark}}, \href{https://doi.org/10.1103/PhysRevD.45.R2188}{\emph{Phys. Rev. D}
  {\bfseries 45} (1992) R2188}.

\bibitem{Damgaard:2019lfh}
P.H.~Damgaard, K.~Haddad and A.~Helset, \emph{{Heavy Black Hole Effective
  Theory}}, \href{https://doi.org/10.1007/JHEP11(2019)070}{\emph{JHEP}
  {\bfseries 11} (2019) 070}
  [\href{https://arxiv.org/abs/1908.10308}{{\ttfamily 1908.10308}}].

\bibitem{Heinonen:2012km}
J.~Heinonen, R.J.~Hill and M.P.~Solon, \emph{{Lorentz invariance in heavy
  particle effective theories}},
  \href{https://doi.org/10.1103/PhysRevD.86.094020}{\emph{Phys. Rev. D}
  {\bfseries 86} (2012) 094020}
  [\href{https://arxiv.org/abs/1208.0601}{{\ttfamily 1208.0601}}].

\bibitem{Parra-Martinez:2020dzs}
J.~Parra-Martinez, M.S.~Ruf and M.~Zeng, \emph{{Extremal black hole scattering
  at $\mathcal{O}(G^3)$: graviton dominance, eikonal exponentiation, and
  differential equations}},
  \href{https://doi.org/10.1007/JHEP11(2020)023}{\emph{JHEP} {\bfseries 11}
  (2020) 023} [\href{https://arxiv.org/abs/2005.04236}{{\ttfamily
  2005.04236}}].

\bibitem{Balasubramanian:2019stt}
V.~Balasubramanian, B.~Craps, M.~De~Clerck and K.~Nguyen, \emph{{Superluminal
  chaos after a quantum quench}},
  \href{https://doi.org/10.1007/JHEP12(2019)132}{\emph{JHEP} {\bfseries 12}
  (2019) 132} [\href{https://arxiv.org/abs/1908.08955}{{\ttfamily
  1908.08955}}].

\bibitem{Kol:2021jjc}
U.~Kol, D.~O'connell and O.~Telem, \emph{{The radial action from probe
  amplitudes to all orders}},
  \href{https://doi.org/10.1007/JHEP03(2022)141}{\emph{JHEP} {\bfseries 03}
  (2022) 141} [\href{https://arxiv.org/abs/2109.12092}{{\ttfamily
  2109.12092}}].

\bibitem{Kim:2023qbl}
S.~Kim, P.~Kraus, R.~Monten and R.M.~Myers, \emph{{S-Matrix Path Integral
  Approach to Symmetries and Soft Theorems}},
  \href{https://arxiv.org/abs/2307.12368}{{\ttfamily 2307.12368}}.

\bibitem{Polchinski:1998rq}
J.~Polchinski, \emph{{String theory. Vol. 1: An introduction to the bosonic
  string}}, Cambridge Monographs on Mathematical Physics, Cambridge University
  Press (12, 2007),
  \href{https://doi.org/10.1017/CBO9780511816079}{10.1017/CBO9780511816079}.

\bibitem{Mogull:2020sak}
G.~Mogull, J.~Plefka and J.~Steinhoff, \emph{{Classical black hole scattering
  from a worldline quantum field theory}},
  \href{https://doi.org/10.1007/JHEP02(2021)048}{\emph{JHEP} {\bfseries 02}
  (2021) 048} [\href{https://arxiv.org/abs/2010.02865}{{\ttfamily
  2010.02865}}].

\bibitem{Porto:2005ac}
R.A.~Porto, \emph{{Post-Newtonian corrections to the motion of spinning bodies
  in NRGR}}, \href{https://doi.org/10.1103/PhysRevD.73.104031}{\emph{Phys. Rev.
  D} {\bfseries 73} (2006) 104031}
  [\href{https://arxiv.org/abs/gr-qc/0511061}{{\ttfamily gr-qc/0511061}}].

\bibitem{Kalin:2020mvi}
G.~K\"alin and R.A.~Porto, \emph{{Post-Minkowskian Effective Field Theory for
  Conservative Binary Dynamics}},
  \href{https://doi.org/10.1007/JHEP11(2020)106}{\emph{JHEP} {\bfseries 11}
  (2020) 106} [\href{https://arxiv.org/abs/2006.01184}{{\ttfamily
  2006.01184}}].

\bibitem{Liu:2021zxr}
Z.~Liu, R.A.~Porto and Z.~Yang, \emph{{Spin Effects in the Effective Field
  Theory Approach to Post-Minkowskian Conservative Dynamics}},
  \href{https://doi.org/10.1007/JHEP06(2021)012}{\emph{JHEP} {\bfseries 06}
  (2021) 012} [\href{https://arxiv.org/abs/2102.10059}{{\ttfamily
  2102.10059}}].

\bibitem{Geroch:1986jjl}
R.P.~Geroch and J.H.~Traschen, \emph{{Strings and Other Distributional Sources
  in General Relativity}},
  \href{https://doi.org/10.1103/PhysRevD.36.1017}{\emph{Conf. Proc. C}
  {\bfseries 861214} (1986) 138}.

\bibitem{Bern:1994zx}
Z.~Bern, L.J.~Dixon, D.C.~Dunbar and D.A.~Kosower, \emph{{One loop n point
  gauge theory amplitudes, unitarity and collinear limits}},
  \href{https://doi.org/10.1016/0550-3213(94)90179-1}{\emph{Nucl. Phys. B}
  {\bfseries 425} (1994) 217}
  [\href{https://arxiv.org/abs/hep-ph/9403226}{{\ttfamily hep-ph/9403226}}].

\bibitem{Bern:1994cg}
Z.~Bern, L.J.~Dixon, D.C.~Dunbar and D.A.~Kosower, \emph{{Fusing gauge theory
  tree amplitudes into loop amplitudes}},
  \href{https://doi.org/10.1016/0550-3213(94)00488-Z}{\emph{Nucl. Phys. B}
  {\bfseries 435} (1995) 59}
  [\href{https://arxiv.org/abs/hep-ph/9409265}{{\ttfamily hep-ph/9409265}}].

\bibitem{Kosmopoulos:2020pcd}
D.~Kosmopoulos, \emph{{Simplifying D-dimensional physical-state sums in gauge
  theory and gravity}},
  \href{https://doi.org/10.1103/PhysRevD.105.056025}{\emph{Phys. Rev. D}
  {\bfseries 105} (2022) 056025}
  [\href{https://arxiv.org/abs/2009.00141}{{\ttfamily 2009.00141}}].

\bibitem{Akhoury:2013yua}
R.~Akhoury, R.~Saotome and G.~Sterman, \emph{{High Energy Scattering in
  Perturbative Quantum Gravity at Next to Leading Power}},
  \href{https://doi.org/10.1103/PhysRevD.103.064036}{\emph{Phys. Rev. D}
  {\bfseries 103} (2021) 064036}
  [\href{https://arxiv.org/abs/1308.5204}{{\ttfamily 1308.5204}}].

\bibitem{Kosower:2018adc}
D.A.~Kosower, B.~Maybee and D.~O'Connell, \emph{{Amplitudes, Observables, and
  Classical Scattering}},
  \href{https://doi.org/10.1007/JHEP02(2019)137}{\emph{JHEP} {\bfseries 02}
  (2019) 137} [\href{https://arxiv.org/abs/1811.10950}{{\ttfamily
  1811.10950}}].

\bibitem{Herrmann:2021lqe}
E.~Herrmann, J.~Parra-Martinez, M.S.~Ruf and M.~Zeng, \emph{{Gravitational
  Bremsstrahlung from Reverse Unitarity}},
  \href{https://doi.org/10.1103/PhysRevLett.126.201602}{\emph{Phys. Rev. Lett.}
  {\bfseries 126} (2021) 201602}
  [\href{https://arxiv.org/abs/2101.07255}{{\ttfamily 2101.07255}}].

\bibitem{Herrmann:2021tct}
E.~Herrmann, J.~Parra-Martinez, M.S.~Ruf and M.~Zeng, \emph{{Radiative
  classical gravitational observables at $ \mathcal{O} $(G$^{3}$) from
  scattering amplitudes}},
  \href{https://doi.org/10.1007/JHEP10(2021)148}{\emph{JHEP} {\bfseries 10}
  (2021) 148} [\href{https://arxiv.org/abs/2104.03957}{{\ttfamily
  2104.03957}}].

\bibitem{Schwinger:1960qe}
J.S.~Schwinger, \emph{{Brownian motion of a quantum oscillator}},
  \href{https://doi.org/10.1063/1.1703727}{\emph{J. Math. Phys.} {\bfseries 2}
  (1961) 407}.

\bibitem{Keldysh:1964ud}
L.V.~Keldysh, \emph{{Diagram technique for nonequilibrium processes}},
  {\emph{Zh. Eksp. Teor. Fiz.} {\bfseries 47} (1964) 1515}.

\bibitem{Calzetta:2008iqa}
E.A.~Calzetta and B.-L.B.~Hu, \emph{{Nonequilibrium Quantum Field Theory}},
  Oxford University Press (2009),
  \href{https://doi.org/10.1017/9781009290036}{10.1017/9781009290036}.

\bibitem{Edison:2022cdu}
A.~Edison and M.~Levi, \emph{{A tale of tails through generalized unitarity}},
  \href{https://arxiv.org/abs/2202.04674}{{\ttfamily 2202.04674}}.

\bibitem{Jakobsen:2022psy}
G.U.~Jakobsen, G.~Mogull, J.~Plefka and B.~Sauer, \emph{{All things retarded:
  radiation-reaction in worldline quantum field theory}},
  \href{https://doi.org/10.1007/JHEP10(2022)128}{\emph{JHEP} {\bfseries 10}
  (2022) 128} [\href{https://arxiv.org/abs/2207.00569}{{\ttfamily
  2207.00569}}].

\bibitem{Jakobsen:2023hig}
G.U.~Jakobsen, G.~Mogull, J.~Plefka and B.~Sauer, \emph{{Dissipative Scattering
  of Spinning Black Holes at Fourth Post-Minkowskian Order}},
  \href{https://doi.org/10.1103/PhysRevLett.131.241402}{\emph{Phys. Rev. Lett.}
  {\bfseries 131} (2023) 241402}
  [\href{https://arxiv.org/abs/2308.11514}{{\ttfamily 2308.11514}}].

\bibitem{Kalin:2022hph}
G.~K\"alin, J.~Neef and R.A.~Porto, \emph{{Radiation-reaction in the Effective
  Field Theory approach to Post-Minkowskian dynamics}},
  \href{https://doi.org/10.1007/JHEP01(2023)140}{\emph{JHEP} {\bfseries 01}
  (2023) 140} [\href{https://arxiv.org/abs/2207.00580}{{\ttfamily
  2207.00580}}].

\end{thebibliography}\endgroup
\bibliographystyle{jhep}
\end{document}